\documentclass[aps,pra,pdf,superscriptaddress,amsmath,amssymb,amsfonts,twocolumn,showpacs,nofootinbib]{revtex4-1}

\usepackage{amssymb}
\usepackage{eqnarray,amsmath}
\newcommand{\abs}[1]{\left| #1 \right|} 

\usepackage{graphicx}
\usepackage{epsfig}
\usepackage{dcolumn}
\usepackage{bm}
\usepackage{braket}
\usepackage{amsmath}
\usepackage{hyperref}
\usepackage{physics}
\usepackage{mathtools}
\usepackage{graphicx,color,xcolor,colortbl}

\usepackage[normalem]{ulem} 

\newcommand{\mc}[2]{\multicolumn{#1}{c}{#2}}
\definecolor{Gray}{gray}{0.85}
\definecolor{LightCyan}{rgb}{0.88,1,1}

\newcolumntype{a}{>{\columncolor{Gray}}c}
\newcolumntype{b}{>{\columncolor{white}}c}

\begin{document}

\title{Control of $^{164}$Dy Bose-Einstein condensate phases\\ and dynamics with dipolar anisotropy}

\author{S. Halder}
\affiliation{Department of Physics, Indian  Institute of Technology Kharagpur, Kharagpur 721302, India} 
\author{K. Mukherjee}
\affiliation{Department of Physics, Indian Institute of Technology Kharagpur, Kharagpur 721302, India}
\affiliation{Mathematical Physics and NanoLund, Lund University, P.O. Box 118, 22100 Lund, Sweden}
\affiliation{Department of Physical Sciences, Indian Institute of Science Education and Research
Kolkata, Mohanpur-741246, West Bengal, India}
\author{S. I. Mistakidis}
\affiliation{ITAMP, Center for Astrophysics $|$ Harvard $\&$ Smithsonian, Cambridge, MA 02138 USA}
\affiliation{Department of Physics, Harvard University, Cambridge, Massachusetts 02138, USA}
\author{S. Das}
\affiliation{Department of Physics, Indian  Institute of Technology Kharagpur, Kharagpur 721302, India} 
\author{P. G. Kevrekidis}
\affiliation{Department of Mathematics and Statistics, University of Massachusetts Amherst, Amherst, MA 01003-4515, USA}	
\author{P. K. Panigrahi}
\affiliation{Department of Physical Sciences, Indian Institute of Science Education and Research
Kolkata, Mohanpur-741246, West Bengal, India}

\author{S. Majumder}
\affiliation{Department of Physics, Indian  Institute of Technology Kharagpur, Kharagpur 721302, India}
\author{H. R. Sadeghpour}
\affiliation{ITAMP, Center for Astrophysics $|$ Harvard $\&$ Smithsonian, Cambridge, MA 02138 USA}

\date{\today}

\begin{abstract}
	
We investigate the quench dynamics of quasi-one and two dimensional dipolar Bose-Einstein condensates (dBEC) of $^{164}$Dy atoms under the influence of a fast rotating magnetic field. 
The magnetic field thus controls both the magnitude and sign of the  dipolar potential. 
We account for quantum fluctuations, critical to formation of exotic quantum droplet and supersolid phases in the extended Gross-Pitaevskii formalism, which includes the so-called Lee-Huang-Yang (LHY) correction. An analytical variational ansatz allows us to obtain the phase diagrams of the superfluid and droplet phases. 
The crossover from the superfluid to the supersolid phase and to single and  droplet arrays is probed
with particle number and dipolar interaction. 
The dipolar strength is tuned by rotating the magnetic field with subsequent effects on phase boundaries. 
Following interaction quenches across the aforementioned phases, we monitor the dynamical formation of supersolid clusters or droplet lattices. 
We include losses due to three-body recombination over the crossover regime, where the three-body recombination rate coefficient scales with the fourth power of the scattering length ($a_s)$ or the dipole length ($a_{dd}$). 
For fixed values of the dimensionless parameter, $\epsilon_{dd} = a_{dd}/ a_s$, tuning the dipolar anisotropy leads to an enhancement of the droplet lifetimes. 
\end{abstract}

\maketitle
	
\section{Introduction}\label{Introduction}

Quantum gases~\cite{Lahaye_2009, Baranov_2012} of atomic species of high spin quantum numbers, such as dysprosium~\cite{Lu_2011,Tang_2015} or erbium~\cite{Aikawa_2012} atoms are ideal candidates for probing quantum fluctuations~\cite{chomaz2022dipolar}.  Particularly, the interplay between long-range anisotropic dipole-dipole (DDI) and contact interactions gives rise to a variety of novel phenomena, including anisotropic superfluidity~\cite{Ticknor, Bismut, Wenzel_2018}, appearance of roton excitations~\cite{Blakie_2012, Chomaz_2018, Natale_2019,  petter_2019, Hertkorn_2021_supersolidity, Schmidt_2021}, formation of self-bound quantum droplets~\cite{Ferrier_2016, schmit2016, Santos_2016,Baillie2017, Santos2016, Macia_2016,Chomaz_2016, Farrier2018} and supersolid states~\cite{Roccuzzo2019, Blakie_2020_tube}. 
The latter exhibit both global phase coherence and periodic density modulations~\cite{Gross_1957, GROSS195857, Boninsegni_2012, Boninsegni2012}, due to the breaking of translation invariance, and are associated with the two low-frequency compressive modes~\cite{Tanzi2019} of the dipolar Bose-Einstein condensate. 
Supersolidity has been widely explored in a series of experiments~\cite{Chomaz_2018, Tanzi_2019, Fabian_2019, Chomaz_2019, Natale_2019,Guo_2019, Ilizhfer2021,  Norcia2021, Hertkorn_2021_supersolidity, Fabian2019}, and theoretically in a number of cold atom settings, ranging from Rydberg systems~\cite{Henkel2012, Cinti_2010}, lattice trapped atomic mixtures~\cite{Batrouni_2008, Dang_2008, Suthar_2020, Heidarian_2010}, and with condensates with spin-orbit coupling~\cite{Li_2017, Sachdeva_2020} or coupled to a light field~\cite{Leonard_2017}. 
Vorticity patterns of rotating supersolid dipolar gases have been also identified~\cite{Roccuzzo_2020, Galleimi_2020, Tengstrand_2021, prasad_2019_pra}. 

The existence of distinct phases in dBECs~\cite{Santos2003} is inherently related to the presence of roton excitations~\cite{Landau_1941,Landau_1949, Feynman_1954, GLyde1993}. 
Specifically, a roton minimum appears in the spectrum due to attractive DDI. 
This minimum softens for strong DDI such that the excitation energy tends to zero and the dipolar gas may suffer collapse. 
Quantum fluctuations stabilize the dipolar gas~\cite{chomaz2022dipolar,mistakidis2022cold} balancing against the attractive DDI~\cite{B_ttcher_2021}. 
To first order, they are commonly described by the LHY interaction energy~\cite{Lee1957,Lima_2012}. 
The inclusion of the LHY corrections~\cite{saito_2016, Bisset_2016_GS_phase} leads to an extended Gross-Pitaevskii equation (eGPE)~\cite{Santos_2016,Santos2016,Chomaz_2016,baillie2016self}. 
This treatment can accommodate supersolids as well as single or multiple droplet patterns~\cite{Santos_2016,Santos2016,schmit2016}, whose arrangement depends crucially on the transverse direction~\cite{Bisset_2016_GS_phase,poli2021maintaining}.     

In the majority of theoretical investigations to date, the magnetic field which modifies the contact interaction remains fixed, ~\cite{Chomaz_2016, Chomaz_2018, Chomaz_2019}. 
In dipolar gases, the possibility exists
(and has remained largely unexplored to the best of our
knowledge) to apply rotating magnetic field with frequency $\Omega$, to control the DDI~\cite{PhysRevLett.89.130401,Baillie_2020, Prasad_2019} through the inherent anisotropy of dipolar interaction.  
When $\Omega$ is smaller (greater) than the Larmor (trap) frequency, the dipoles follow the external field, a process already realized in experiment~\cite{PhysRevLett.120.230401}.
In this regime, {typical} dynamical instabilities triggered by the rotation at lower or of the order of the trap frequency~\cite{Prasad_2019} are suppressed ~\cite{Baillie_2020}. 
As such, it is thus possible to employ the time-averaged DDI, depending on the angle between the dipole and the field axis, to tune the dipole strength and sign. 
It then becomes feasible to enter different phases of the dBEC via adjusting the DDI, to control the interval of existence of the emergent phases, exploit the anisotropy to design dipolar configurations (such as, e.g., specific lattice arrangements), and even suppress the droplet evaporation.

An even less pursued direction is to monitor the dynamical generation of the self-bound states, traversing the relevant phase boundaries, a scenario that has been exploited in experiments~\cite{Farrier2018, Chomaz_2018}. 
It would be interesting to explore differences between the quenched states in the long-time dynamics as compared to the respective ground-state configurations. 
Also, an understanding of the metastable states at intermediate time scales due to specific instabilities and the behavior of the global phase coherence~\cite{Tanzi_2019} is still far from complete. 
Motivated by the intensive experimental and theoretical activity, see for instance the reviews~\cite{Guo_2021, Luo_2017}, we investigate the ground-state phase diagram and quench dynamics of a dBEC under the influence of a rotating magnetic field in both the quasi-1D and the quasi-2D regimes.

We extract the ground-state phase diagram of the quasi-2D dBEC as a function of 
the ratio of dipolar to contact interactions, as
well as of the atom number. 
The emergent phases include the superfluid (SF), the supersolid (SS), and multiple ($\rm DL_{M}$) and single ($\rm DL_{S}$) droplet states. 
It is shown that SS states are characterized by spatially overlapping density humps while droplets form crystalline patterns arranged as lattices with polygonal characteristics~\cite{adhikari_2022}. 
These phases have been shown to occur in fixed magnetic fields~\cite{Tanzi_2019, Chomaz_2018}.
Herein, we determine the explicit boundaries between the different phases, such as the $\rm DL_{M}$ (also known as insulating droplet region~\cite{Chomaz_2019}) and $\rm DL_{S}$ by exploiting the anisotropic nature of the dipole-dipole interaction.
Interestingly, we find that a tilted magnetic field 
favors (independently of the $s$-wave scattering) transitions between the different phases and in particular for angles larger than the magic angle, solely the SF and the $\rm DL_{S}$ persist. The rotating magnetic field alters the configuration of the dBEC, enforcing for instance, broader 2D distributions across the $x$-$y$ plane or, e.g., square and honeycomb lattice structures for angles smaller than the magic angle. 

The interaction quench of a SF state across the relevant phase boundaries, results in the dynamical nucleation of elongated arrays in quasi-1D as also observed in~\cite{B_ttcher_2021, Chomaz_2019} and lattices in quasi-2D of SS and droplets due to the growth of the roton instability~\cite{Hertkorn_2021_supersolidity}. The latter manifests as ring excitations or elliptic halos in the early times before developing into clusters that then saturate. 
Phase coherence is not maintained in the course of the evolution and it is fully lost in the droplet regime. 
Quenches from the SF to the $\rm DL_{S}$ 
phase produce $\rm DL_{M}$ lattices. 

We demonstrate that the number of droplets contained in a lattice is larger for reduced post-quench contact interactions or tilted fields with an angle smaller than the magic angle. 
Also, the amount of dynamically nucleated droplets in the long-time quench dynamics is larger as compared to the respective ground-state post-quench configuration. 
Another central feature of our findings is the exploration of the self-evaporation of the above-discussed structures by including three-body recombination processes into our analysis. 
This mechanism prevails for bound states and raises a nontrivial obstacle 
in connection with the realization of droplets and especially SS phases~\cite{Chomaz_2019,Chomaz_2016}. Specifically, we showcase that the anisotropic magnetic field, lying below the magic angle, is a tool to increase the lifetime of self-bound states. These regions were inaccessible in previous studies~\cite{Ferrier_2016, schmit2016} due to the assumption of an aligned magnetic field along the $z$-direction. 

This work is structured as follows. Section~\ref{theoy_sec} describes the anisotropic dipolar potential and introduces the eGPE framework. In section~\ref{GS_sub}, we extract the ground-state phase diagram of the quasi-2D dBEC. 
The dynamical generation of self-bound SS and droplet states following interaction quenches is discussed in Sec.~\ref{Quench_sub}. 
In Sec.~\ref{TBI_append} we monitor the self-evaporation of the quenched states by accounting for three-body recombination processes. 
A summary of our findings together with future perspectives are provided in Sec.~\ref{conclusion}. Appendix~\ref{VarCal_sub} is devoted to the construction of the variational approach for confirming the existence of the ground-state phases, while Appendix~\ref{numerics} delineates the ingredients of our numerical simulations. 
In Appendix~\ref{breathing}, we briefly analyze the collective excitation processes of the quasi-2D dBEC.

\section{Beyond mean-field treatment of the dipolar condensate}\label{theoy_sec}

Below, we describe the explicit form and properties of the considered DDI potential as well as provide the intrinsic system parameters which closely follow recent experimental settings~\cite{Chomaz_2016, Chomaz_2018, Chomaz_2019}. Afterwards, we introduce the eGPE framework that we shall use in order to track the phase diagram and subsequently monitor the quench dynamics of the dipolar condensate.  

\subsection{Modifying dipolar potential with a rotating magnetic field}\label{dipolar_potential} 

We consider a harmonically trapped dBEC in three-dimensions (3D) whose atoms possess 
a magnetic dipole moment $\mu_{m}$. 
The atomic dipoles are polarized by a rotating uniform magnetic field (in the $x$-$y$ plane) of strength $B$, along $\textbf{e}(t) = \cos\phi \textbf{e}_z + \sin\phi(\cos(\Omega t)\textbf{e}_x + \sin(\Omega t)\textbf{e}_y)$ with $\textbf{e}_x$,  $\textbf{e}_y$, $\textbf{e}_z$ being the unit vectors in the $x$, $y$ and $z$ spatial directions, respectively. 
The field rotation frequency $\Omega$ is chosen to be 
$\omega_i \ll \Omega \ll \omega_L=\mu B /\hbar$, where $\omega_i$, $i=x,y,z$ are the trap frequencies, so as to ensure that the dipoles follow the external field. 
Typical angular frequencies are of the order of $\Omega \geq 5 \times 10^2 \omega$, with $\omega$ denoting the radial $\omega \equiv \omega_x=\omega_y$ (elongated, $\omega \equiv \omega_x$) trap frequency of the quasi-2D (quasi-1D) geometry.   

\begin{figure}
\includegraphics[width=0.48\textwidth]{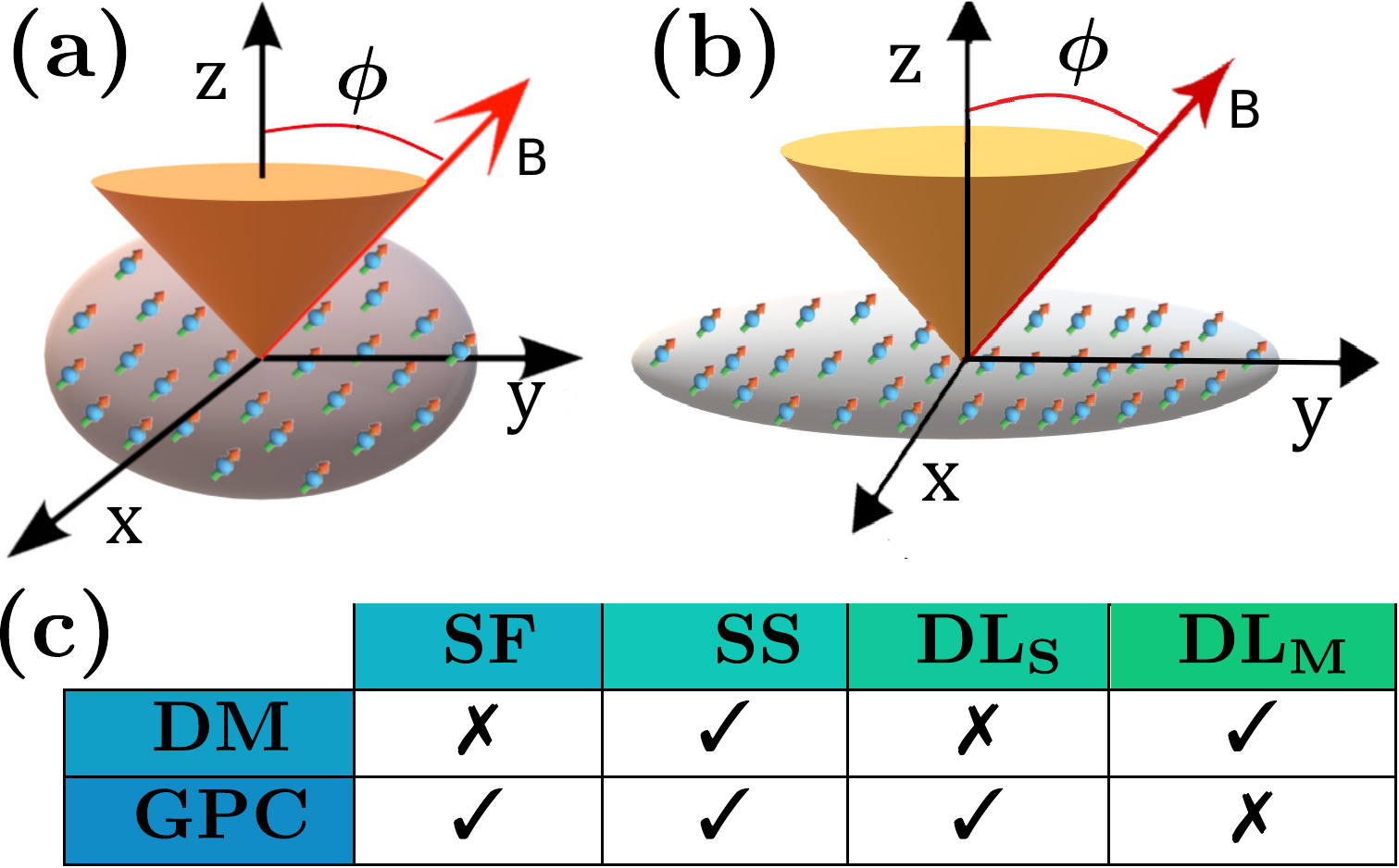}
\caption{Schematic illustration of the dBEC of $^{164}\rm Dy$ atoms trapped in (a)  a circularly-symmetric quasi-2D and (b) an elongated quasi-1D. A magnetic field is applied rotating at an angle $\phi$ around the $z$-axis and aligning the atomic dipoles along its direction. 
(c) The table indicates presence or absence of density modulation (DM) and global phase coherence (GPC) within the superfluid (SF), supersolid (SS), single droplet ($\rm DL_{S}$) and multiple droplet ($\rm DL_{M}$) phases. 
Here we do not refer to typical density modulations originating from the external trap but to the ones stemming from the competition between short- and long-range interactions leading to spatially periodic density undulations. }
\label{Schematic}
\end{figure}

The tilt angle with respect to the $z$-axis is $\phi$, see Fig.~\ref{Schematic}, such that the DDI \cite{Baillie_2020} is given by  $U_{dd} (\textbf{r},t) = \frac{\mu_0 \mu^2_{m}}{4\pi} \left[\frac{1-3(\textbf{e}(t)\cdot\hat{r})^2}{r^3}\right]$,
where $\mu_0$ is the permeability of the vacuum. For $\Omega = 0$ and $\textbf{e}_z \cdot \hat{r} =1$, a head-to-tail arrangement of the dipoles occurs leading to an attractive DDI, i.e. $U_{dd} < 0$. Where $\textbf{e}_z \cdot \hat{r} = 0$, the dipoles are located side-by-side and interact repulsively. 

\begin{figure*}
\includegraphics[width=1.0\textwidth]{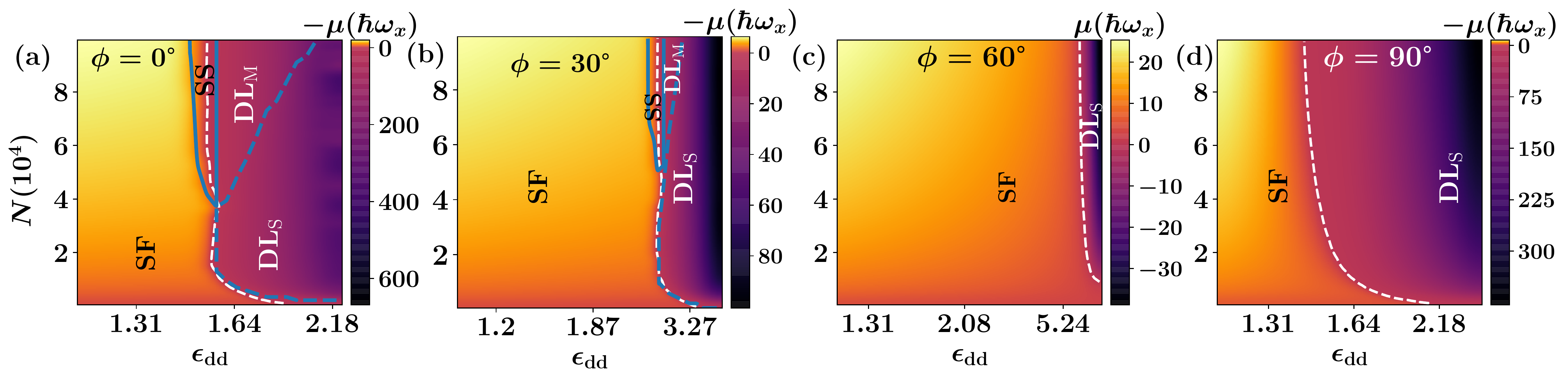}
\caption{Phases identified through the chemical potential of a circular quasi-2D dBEC in terms of the relative interaction strength parameter $\epsilon_{dd}=a_{dd}/a_s$ and the atom number for different magnetic field orientations, namely ($\text{a}$) $\phi = 0^{\circ}$, ($\text{b}$) $\phi = 30^{\circ}$, ($\text{c}$) $\phi = 60^{\circ}$ and ($\text{d}$) $\phi=90^{\circ}$. 
Apparently, as the magnetic field tends towards the $x$-$y$ plane ($\phi=90^{\circ}$) the $\rm DL_{M}$ and SS phases do not occur and solely the SF and $\rm DL_{S}$ survive. 
The white dashed line indicates $\mu=0$, while the blue solid (dashed) line delimits the SS ($\rm DL_{S}$) phase. 
The dBEC resides in a quasi-2D trap characterized by $(\omega_x, \omega_y, \omega_z) = 2\pi \times (45, 45, 133) \rm Hz$.}  
\label{phae_diag_pan}
\end{figure*}

The corresponding time-averaged DDI, over a full rotation cycle of the polarizing magnetic field, is 
\begin{equation}\label{avgddi}
\begin{split}
\langle{U_{dd} (\textbf{r})}\rangle &= \frac{\Omega}{2\pi}\int_{0}^{\frac{2\pi}{\Omega}} U_{dd} (\textbf{r},t) dt 
\\&=\frac{\mu_0 \mu_{m}^2}{4\pi} \left[\frac{1-3(\textbf{e}_z\cdot\hat{r})^2}{r^3}\right]\left(\frac{3 \cos^2 \phi-1}{2}\right).
\end{split}
\end{equation}
Notice that the last factor in Eq.~\eqref{avgddi} decreases from 1 to $-1/2$ when $0<\phi< \pi/2$, and vanishes if $\phi$ equals the magic angle \cite{Baillie_2020} $\phi_m=\cos^{-1}{1/\sqrt{3}}\approx 54.7^{\circ}$. 
The inverted configuration takes place for $\phi>\phi_m$ in which the time-averaged DDI is attractive even though particles reside side-by-side (also known as anti-dipolar regime~\cite{Giovanni2002, Baillie_2020, Prasad_2019}). 
Therefore, the dBEC subjected to this rotating magnetic field effectively experiences the time-averaged DDI potential whose strength and sign can be tuned by varying the tilt angle $\phi$. 
Such a rotating long-range potential has already been implemented~\cite{Tang_2015}, while it has also been theoretically employed in the works of ~\cite{Baillie_2020,prasad_2019_pra}. 
Moreover, we have confirmed within our numerical simulations that this time-averaged consideration does not affect our findings if we compare them to the case where the time-dependent DDI is instantaneously followed~\footnote{The instantaneous DDI in momentum space reads 
$U_{dd}(k,t)=(\mu_{0} \mu^2_{m}/3)\Big[3\Big(k_x \cos(\Omega t)\sin(\phi)+k_y \sin(\Omega t)\sin(\phi)+k_z \cos(\phi)\Big)^2/k^2-1\Big]$.}. 
It is worthwhile to mention that the DDI can be tuned this way independently of the tuning of the zero-range interaction with Feshbach resonance techniques. 
In this manner, it is possible to realize the different dBEC phases by adjusting solely the strength of the DDI, see, e.g., 
Fig.~\ref{phae_diag_pan}. 

Finally, we remark that the presence of rotation should facilitate the tunability of the DDI, especially in the experiment. 
Otherwise, only a tilted magnetic field (in the $x-z$ plane) is characterized by two different angles\footnote{The DDI for a fixed magnetic field ($\Omega=0$) reads in the $x$-$z$ plane reads $U_{dd} = \frac{\mu_0 \mu_{m}^2}{4\pi} \frac{1 - 3 [(\cos(\phi) \textbf{e}_z. \hat{r} + \sin(\phi) \textbf{e}_x .\hat{r} )]^2}{r^3}$.}.

\subsection{Extended Gross-Pitaevskii framework}\label{egpe_sub}

In the ultracold regime the gas is characterized by a single macroscopic wave function, $\psi (\textbf{r} , t) = \langle \hat{\psi}(\vb{r}, t \rangle)$, whose temporal evolution is described by a suitable eGPE~\cite{Lima_2011,Santos_2016, Chomaz_2016, Bisset_2016_GS_phase}. 
The latter incorporates quantum fluctuations in terms of the first order beyond mean-field LHY  correction contribution and in particular reads
\begin{eqnarray}\label{eGPE}   
& i\hbar \frac{\partial \psi(\textbf{r},t)}{\partial t}  =  \bigg[-\frac{\hbar^2}{2m}\nabla^2 + V(\textbf{r}) + g \abs{\psi(\textbf{r},t)}^2+\nonumber\\& \gamma(\epsilon_{dd})\abs{\psi(\textbf{r},t)}^3 +\int dr^{\prime} U_{dd}(\textbf{r-r}^{\prime})\abs{\psi(\textbf{r}^{\prime},t)}^2 \bigg] \psi(\vb{r},t). 
\end{eqnarray}
Here, the 3D harmonic trap is $V(\textbf{r}) = m(\omega_x^2 x^2 + \omega_y^2 y^2 + \omega_z^2 z^2)/2$ and $m$ is the atom mass. 
Apart from the long-range time-averaged DDI in Eq.~\eqref{avgddi}, the atoms collide via short-range contact potential, characterized by the effective strength $g=4\pi\hbar^2 a_s/ m$, with $a_s$ being the 3D $s$-wave scattering length. 
The penultimate term in Eq.~\eqref{eGPE} denotes the LHY contribution which is crucial for the realization of many-body self-bound states such as the $\rm DL_{s}$ or $\rm DL_{M}$ dipolar droplets, as well as the SS phase. 
We remark that in 3D, quantum fluctuations scale with the gas density as $\sim n^{3/2}$. 
In a harmonic trap, the LHY correction can be incorporated in the eGPE, with the local density approximation, i. e. $n\to n(r,t)=\abs{\psi(\textbf{r},t)}^2$, and with  $\gamma(\epsilon_{dd}) = \frac{32}{3}g \sqrt{\frac{a_s^3}{\pi}} \left(1+\frac{3}{2}\epsilon_{dd}^2\right)$~\cite{Lima_2011, Lima_2012}. 
Importantly, the dimensionless parameter $\epsilon_{dd} = a_{dd}/a_s$ with  $a_{dd}= \mu_0 \mu^2_{m} m /12\pi\hbar^2$ being the dipolar length, quantifies the relative strength of the DDI as compared to the contact interaction. 

Below, we shall reveal the emergent ground-state phases of the dBEC stemming from the interplay between the dipolar and contact interactions, employing the parameters of the recent experiments, but now accounting for a rotating magnetic field~\cite{Chomaz_2018,Chomaz_2019, Natale_2019}. 
Subsequently, the dynamical deformation of the identified dipolar configurations is monitored upon considering quenches of the $s$-wave scattering length, and thus of $\epsilon_{dd}$, across the aforementioned phases. 
A particular emphasis is placed on the role of dimensionality ranging from (i) an elongated quasi-1D trap with frequencies $(\omega_x,\omega_y,\omega_z)= 2\pi \times (227,37,135) \rm Hz$~\cite{Chomaz_2019} to (ii) a circularly symmetric quasi-2D trap characterized by $(\omega_x,\omega_y,\omega_z)= 2\pi \times (45,45,133) \rm Hz$~\cite{Bisset_2016_GS_phase}, see also Fig.~\ref{Schematic}($\text{a}$), ($\text{b}$). 
Our results can be replicated using a dBEC of $^{164}\rm Dy$ atoms having magnetic moment $\mu_{m}=9.93\mu_{B}$, where $\mu_{B}$ is the Bohr magneton. 
The dipolar length is $a_{dd} =131a_{B}$, where $a_{B}$ is the Bohr radius. 

The characteristic timescale is set by $\omega_x^{-1}=3.5$  ms ($\omega_x^{-1}=0.7$  ms) in the quasi-2D (quasi-1D) case. 
Similarly the respective length scale refers to the harmonic oscillator length being $l_{{\rm osc}}=\sqrt{\hbar/(m \omega_x)}= 1.17 \mu m$ ($ 0.52 \mu m$) for the quasi-2D (quasi-1D) setting.

\section{Phase diagram of \text{d}BECs in rotating magnetic fields}\label{GS_sub}

In the following, we investigate the ground-state phases of the 2D dBEC arising for different atom numbers and relative interactions  $\epsilon_{dd}=a_{dd}/a_s$. 
A central aim of our discussion is to unravel the role of the orientation of the dipoles, as dictated by the titled time-averaged magnetic field [Fig.~\ref{Schematic}($\text{b}$)], on the emergent structural configurations. 
The latter are identified through the relevant integrated density profiles $n(x,y)=\int dz~n(x,y,z,t)$ and $n(y,z)=\int dx~n(x, y,z,t)$ being experimentally detectable e.g. via {\it in-situ} imaging~\cite{chomaz2022dipolar,Sohmen_2021} and herein are normalized to the particle number. 

The circularly symmetric quasi-2D trap geometry is realized by applying a tight confinement in the transversal $z$-direction [Fig.~\ref{Schematic}(a)]. 
The same structures occur also in quasi-1D [Fig.~\ref{Schematic}($\text{b}$)], but are omitted here for brevity. 
For completeness, in Appendix~\ref{VarCal_sub}, we benchmark the properties of the static (SF and ${\rm DL_{S}}$) phases found within the eGPE approach utilizing a variational ansatz. 
A similar approach has also been recently leveraged within the dBEC context e.g. in Refs.~\cite{Goral_2021,pal2020excitations}. 
All ground states are obtained by propagating Eq.~(\ref{eGPE}) in imaginary time with a split-step Crank-Nicolson approach (Appendix~\ref{numerics}).

\subsection{Aligned dipolar BEC}\label{Quasi2D_sub}

To distinguish the various dBEC phases, we employ as a ``control" parameter the chemical potential related to the total energy $E$ (see Eq.~(\ref{eq:energy_dBEC})) via $\mu =\partial E[\psi]/\partial n$. 
Naturally, a SF state has $\mu>0$, whereas the droplet configurations occur for $\mu < 0$ since they refer to self-bound states. 
In addition we will show that the SS phase appears in the vicinity of $\mu \rightarrow 0$.

\begin{figure}
\begin{center}
\includegraphics[width=0.49\textwidth]{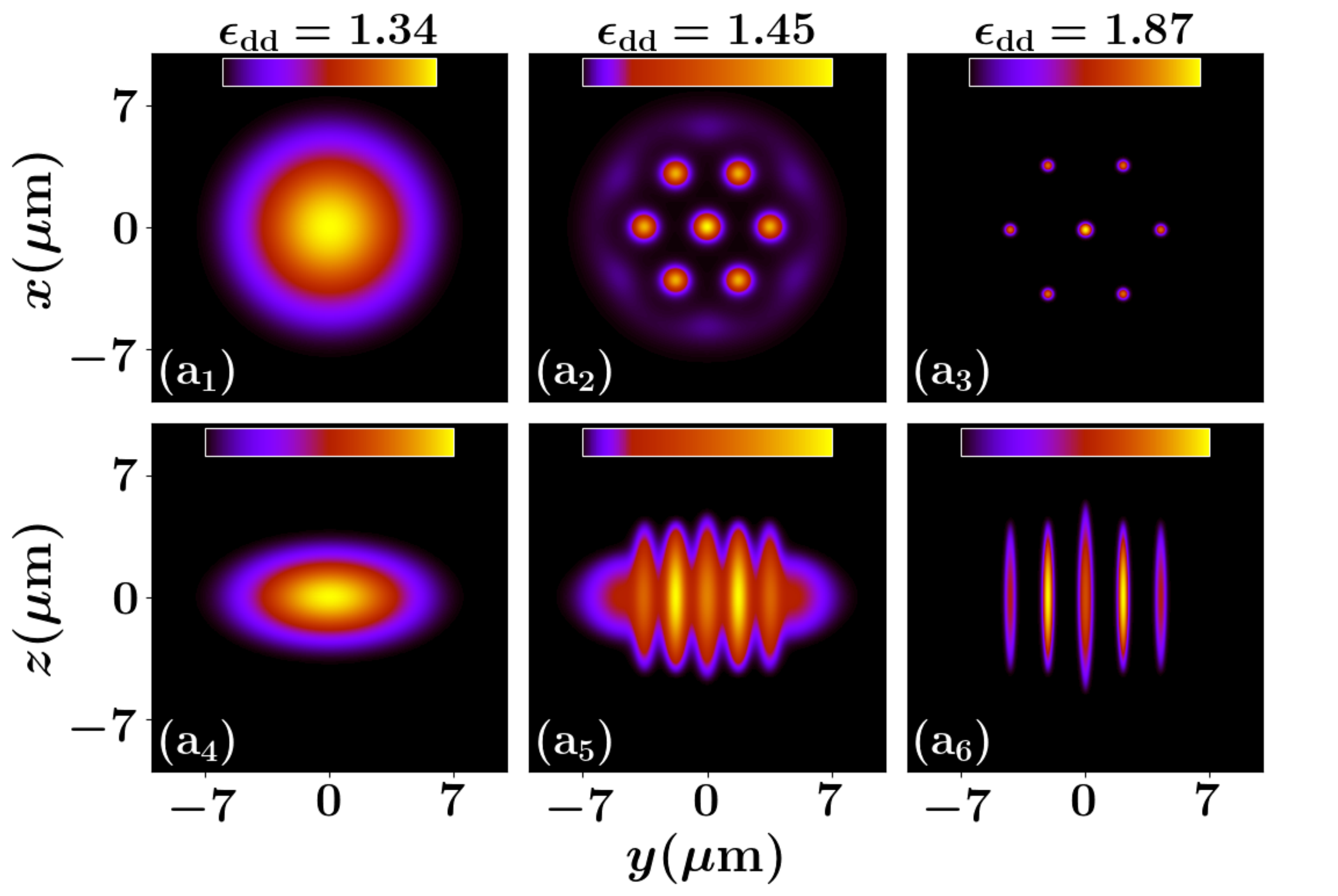}
\caption{Ground-state density profiles ($\text{a}_1$)-($\text{a}_3$) $n(x,y)$ and ($\text{a}_4$)-($\text{a}_6$) $n(y,z)$ representing ($\text{a}_1$), ($\text{a}_4$) a SF, ($\text{a}_2$), ($\text{a}_5$) a SS and ($\text{a}_3$), ($\text{a}_6$) a droplet (hexagonal) lattice. 
A SS state is characterized by overlapping density humps, while a droplet cluster has a crystal arrangement. 
The harmonically trapped quasi-2D dBEC with $(\omega_x, \omega_y, \omega_z)= 2\pi \times (45, 45, 133)\rm Hz$ has $N=10^{5}$ particles and it is subjected to a magnetic field along the $z$-direction, i.e. $\phi=0^{\circ}$. 
The colorbar alternates for each panel and corresponds to the density changing from zero (black) to a maximum value (yellow) in units of $1/l_{{\rm osc}}^2=0.73 \mu m^{-2}$}. 
\label{fig_quasi_2d_den1}
\end{center}
\end{figure}

At $\phi = 0^{\circ}$, where the external field forces the dipoles to be oriented along the $z$-axis, the phase arising at small magnitudes of $\epsilon_{dd}$ (e.g. $\epsilon_{dd} < 1.45$ for $N=60000$) has $\mu>0$, see Fig.~\ref{phae_diag_pan}($\text{a}$). As such, a typical SF state emerges characterized by a smooth 2D Thomas-Fermi (TF) density distribution along the $x$-$y$ plane, while being compressed along the $z$-axis due to the tight confinement, see $n(x,y)$ and $n(y,z)$ in Fig.~\ref{fig_quasi_2d_den1}($\text{a}_1$) and Fig.~\ref{fig_quasi_2d_den1}($\text{a}_4$). 
This SF phase boundary ($\epsilon_{dd}<1.4$)
has been reported for harmonically trapped 3D dBECs subjected to a static magnetic field~\cite{Galleimi_2020}.

However, for $\epsilon_{dd}$ above a certain critical value, indicated by the dashed white line in Fig.~\ref{phae_diag_pan}($\text{a}$), the system transitions to a negative $\mu$ region. 
The latter regime accommodates distinct phases of matter that occur due to the existence of quantum fluctuations~\cite{Tanzi_2019}. Indeed, with increasing $\epsilon_{dd}$ for large atom number ($N>2 \times10^4$) and in the vicinity of $\mu \rightarrow 0$ (see the bounded by solid blue line area in Fig.~\ref{phae_diag_pan}($\text{a}$)), a periodic density modulated pattern develops along the weakly confined $x$-$y$ plane. Particularly, a hexagonal lattice appears having  its seven density humps inter-linked by lower density regions [see $n(x,y)$ in  Fig.~\ref{fig_quasi_2d_den1}($\text{a}_2$)]. 
This relatively small spatial overlap establishes a global phase coherence~\cite{Roccuzzo2019,Ilizhfer2021,Chomaz_2019} across the dBEC [Fig.~\ref{Schematic}(c)]. 
The aforementioned individual density humps are also imprinted in the transversal $z$-direction with $n(y,z)$ featuring fringes at their location [Fig.~\ref{fig_quasi_2d_den1}($\text{a}_5$)]. 
Such a SS phase~\cite{Fabian2019, Chomaz_2019,Sohmen_2021} has been recently identified theoretically~\cite{poli2021maintaining,adhikari_2022} and realized experimentally~\cite{Bland2022}.  

A further increase of $\epsilon_{dd}$ 
(smaller $a_s$) results in a dramatic suppression of the density overlap among the individual density humps [Fig.~\ref{fig_quasi_2d_den1}($\text{a}_3$)]. This behavior occurs for large atom numbers and large negative $\mu$ with the system entering the $\rm DL_{M}$ phase [Fig.~\ref{phae_diag_pan}($\text{a}$)]. 
A typical example is provided in Fig.~\ref{fig_quasi_2d_den1}($\text{a}_3$) where a hexagonal crystal structure builds-upon the $x$-$y$ plane with isolated stripe patterns being evident in $n(y,z)$ [Fig.~\ref{fig_quasi_2d_den1}($\text{a}_6$)]. 
With smaller atom numbers, the dBEC transits to the $\rm DL_{S}$ phase. 
It is also evident that for $N<4 \times 10^4$ where finite size effects are expected to play a crucial role~\cite{Kwon2021}, the system deforms from a SF to a $\rm DL_{S}$ state and vice versa with tuning $\epsilon_{dd}$. 
Notice that the $\rm DL_{S}$ and SF phases, 
are characterized by a zero global phase coherence [see also Fig.~\ref{Schematic}($\text{c}$)] but differ substantially in their spatial localization and importantly the former is self-bound ($\mu<0$).

\subsection{Anisotropic dipolar BEC}

Next, we investigate the impact of a rotating magnetic field on the emergent phase diagram with varying contact interaction and atom number in Fig.~\ref{phae_diag_pan}($\text{b}$)-($\text{d}$). 
At the magic angle $\phi_m$, where $\langle U_{{\rm dd}} \rangle=0$, only SF states form, independently of the value of the $\epsilon_{dd}$ (not shown). 
In fact, as $\phi\rightarrow \phi_m$, the overall dipole interaction strength decreases and the contact interaction dominates favoring SF formation. 
Notice the extended SF phase for $\mu>0$ and $\phi = 30^{\circ}$ illustrated in Fig.~\ref{phae_diag_pan}($\text{b}$). 
Accordingly, the interaction (or $\epsilon_{dd}$) intervals within which modulated density structures form (i.e. either SS or $\rm {DL}_{M}$ states) shrink and are shifted towards larger $\epsilon_{dd}$. 
This shift (marked by dashed white lines) occurs consistently as long as $\phi< \phi_m$ and all four different phases can be realized. 
Here, tuning $\phi$ to larger values within the $\rm DL_{M}$ phase produces larger droplet lattices eventually transitioning to a SS (see also Fig.~\ref{fig_phi_den}($\text{b}_1$)-($\text{b}_3$) and the discussion below), while leaving a SF state almost unaltered only affecting its spatial width. 
Note in passing that the aforementioned shrinkage of the $\rm DL_{M}$ phase as $\phi \to \phi_m$ appears also in quasi-1D (not shown). 
However, it is more prominent in the quasi-2D setting revealing that here for $\phi \to \phi_m$ the isolated droplets are more prone to coalesce into a single droplet configuration. 
The difference between quasi-1D and quasi-2D is traced back to the fact that in the latter setup, and for our parameter set, the magnitude of the attractive DDI is enhanced.

When $\phi>\phi_{m}$, the dipoles become progressively more attractive to each other in the $x$-$y$ plane, see Fig.~\ref{phae_diag_pan}($\text{c}$) and ($\text{d}$), and the self-bound DL$_S$ region extends towards smaller $\epsilon_{dd}$. 
We note that while $\mu$ is negative below a critical $\epsilon_{dd}$ (see the dashed white line in Fig.~\ref{phae_diag_pan}($\text{c}$)-($\text{d}$)), unlike in the $\phi<\phi_m$ scenario, there are no states characterized by a density modulation along the $x$-$y$ plane. 
This again is a consequence of the strongly attractive DDI preventing the emergence of density undulated patterns. 
In this way, both the $\rm DL_{M}$ and SS phases disappear and only the $\rm DL_{S}$ ($\mu < 0$) and SF ($\mu >0$) states occur.

\begin{figure}
\includegraphics[width=0.48\textwidth]{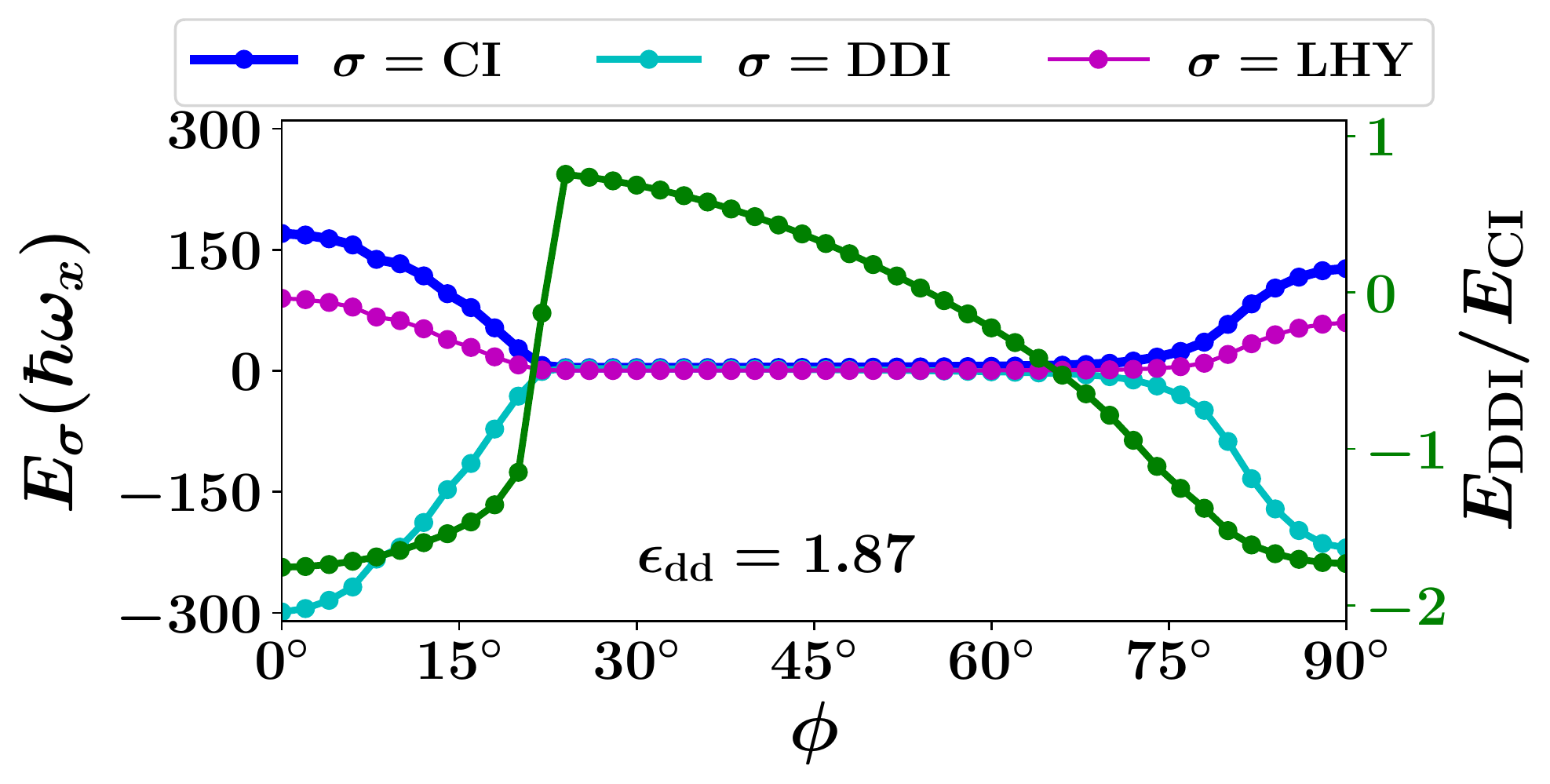}
\caption{Variation of the different energy components $E_{\sigma}$ (see the legend) as a function of the tilt angle $\phi$ at a fixed $\epsilon_{dd}=1.87$ for the quasi-2D dBEC with $(\omega_x, \omega_y, \omega_z) = 2 \pi \times (45, 45, 133)$Hz.   
At a fixed $\epsilon_{dd}=1.87$, droplets are generated within the $\phi<22^{\circ}$ and $\phi>\phi_m$ regions where $E_{\rm{DDI}}<0$. The green line refers to $E_{\rm{DDI}}/E_{\rm{CI}}$ (right axis) with CI denoting the contact interaction.}
\label{fig:Energy_gs}
\end{figure}

The total energy stored in a dBEC can be organized as follows
\begin{equation}
E = E_{\rm K} + E_{\rm V} + E_{\rm CI} + E_{\rm DDI} + E_{\rm LHY}.\label{eq:energy_dBEC}
\end{equation}
In this expression, the energy contributions $E_{\rm CI}$, $E_{\rm DDI}$, $E_{\rm LHY}$ refer to the contact, the dipolar, and the beyond mean-field LHY interaction energies respectively. 
These energy terms dictate the generation of the different phases. 
Also, $E_{\rm K}$ denotes the kinetic energy and $E_{\rm V}$ is the external potential energy.

The dependence of $E_{\rm CI}$, $E_{\rm DDI}$ and $E_{\rm LHY}$ with $\phi$ for $\epsilon_{dd}=1.87$ and $N=10^{5}$ is shown in Fig.~\ref{fig:Energy_gs}. 
Evidently, $E_{\rm CI}$, $E_{\rm DDI}$, $E_{\rm LHY}$ are positive in the interval $22^{\circ} < \phi < \phi_{m}$, and  thus a SF is formed. 
Particularly, $E_{\rm CI}$ is dominant and all other contributions are considerably weaker. 
However, in the case of $\phi = 22^{\circ}$, $E_{\rm DDI}$ is negative and gradually drops as $\phi$ lowers. 
A similar behavior of the energy constituents is observed for $\phi>\phi_m$ where the $\rm DL_{S}$ state appears. 
Therefore in these two regions, namely $\phi < 22^{\circ}$ and $\phi>\phi_m$ the rapid increase of $\abs{E_{\rm DDI}}$ towards negative values, driving the dBEC to collapse, is actually compensated by the enhanced combined repulsive contribution of $E_{\rm LHY}$ and $E_{\rm CI}$ occurring for smaller (larger) $\phi$ in the first (second) region. 
This competition leads to the formation of stable droplet states  in the corresponding $\phi$ intervals. 
Such a stabilization mechanism occurs also for the case of an aligned field ($\phi=0$) in terms of $\epsilon_{dd}$~\cite{chomaz2022dipolar}.

\begin{figure}
\begin{center}
\includegraphics[width=0.49\textwidth]{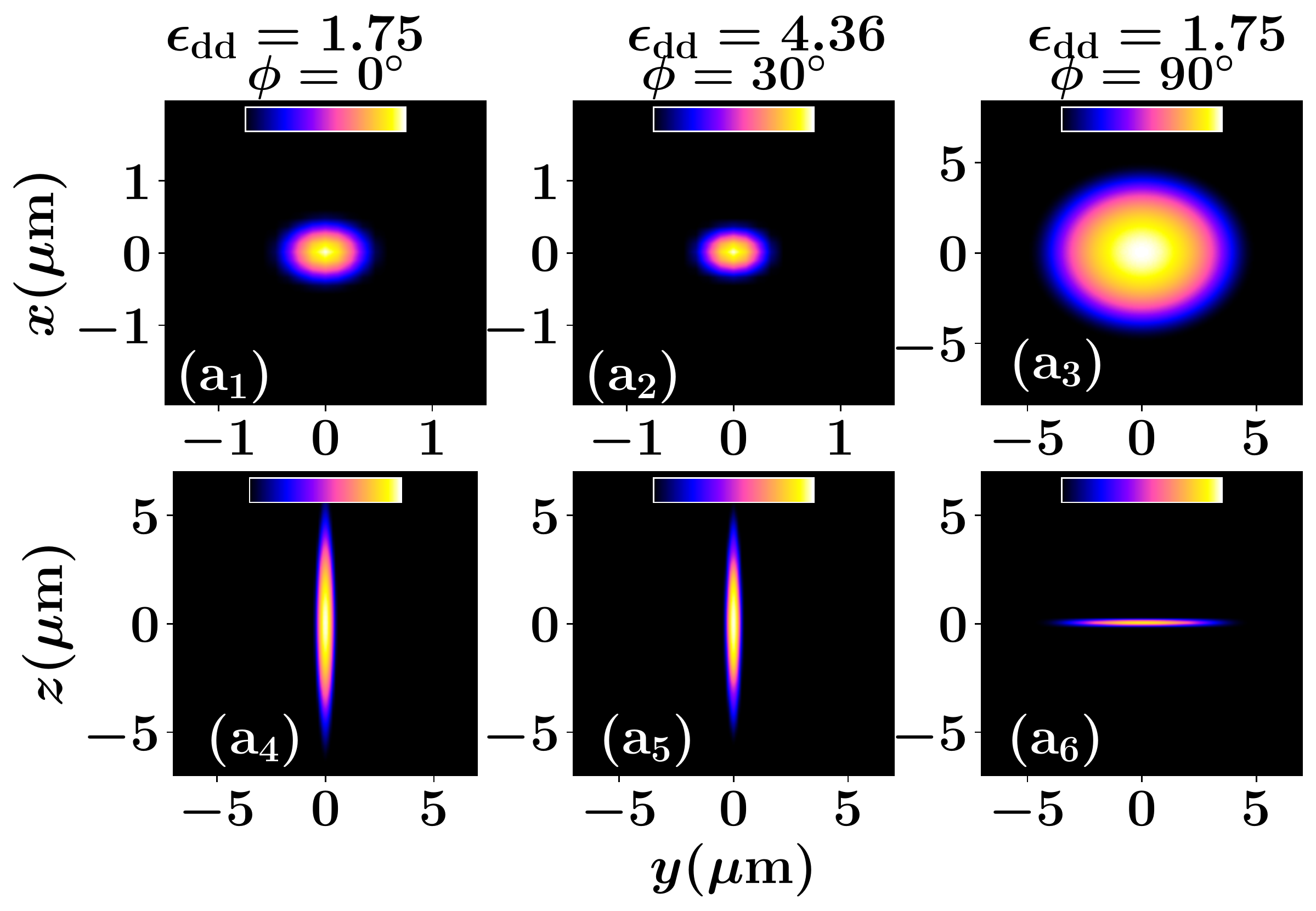}
\caption{Impact of the field orientation $\phi$ on the spatial distribution of the $\rm DL_{S}$ state. Densities ($\text{a}_1$)-($\text{a}_3$) $n(x,y)$ and ($\text{a}_4$)-($\text{a}_6$) $n(y,z)$ of a dBEC with $N=6\times 10^{4}$ in a quasi-2D trap of $(\omega_x,\omega_y,\omega_z) = 2\pi \times (45, 45, 133)\rm Hz$. 
The single droplet becomes narrower (wider) for larger $\phi$ in the interval $\phi<\phi_m$ ($\phi>\phi_m$) and becomes elongated along the $z$- ($y$-) direction due to the combined effect of the polarization of the magnetic field and $\epsilon_{dd}$. 
The characteristic length scale set by the trap is $l_{{\rm osc}}=1.17 \mu m$. 
The colorbar expressed in units of $1/l_{{\rm osc}}^2=0.73 \mu m^{-2}$ is different for each panel denotes the density with intensity from zero (black) to a maximum value (yellow).}
\label{fig_quasi_2d_den2}
\end{center}
\end{figure} 

The spatial distribution of DL$_S$ is shown in Fig.~\ref{fig_quasi_2d_den2} at several values of $\phi$ and $\epsilon_{dd}$; $\epsilon_{dd}$ is varied so that we remain in the same phase. 
Recall that the $\rm DL_{S}$ is shifted to larger $\epsilon_{dd}$ as $\phi \to \phi_m$ [Fig.~\ref{phae_diag_pan}] due to the accompanied weakening of the effective dipole interaction [Eq.~\eqref{avgddi}]. 
{As such, we choose to present the $\rm DL_{S}$ state in Fig.~\ref{fig_quasi_2d_den2}($\text{a}_2$) at $\epsilon_{dd}=4.36$ for $\phi=30^{\circ}$.}
The larger $\epsilon_{dd}$ (smaller $a_s$) is responsible for the decreasing width of $n(x,y)$ when $\phi \to \phi_m$ [Fig.~\ref{fig_quasi_2d_den2}$(\text{a}_1)$, ($\text{a}_2$)]. 
However, as the anisotropy is changed through $\phi$ with $\phi<\phi_m$ and fixed $\epsilon_{dd}$ the dipoles become less repulsive along the $x$-$y$ plane enforcing a spreading of $n(x,y)$ (not shown). 
A further increase of $\phi$ beyond $\phi_m$ leads to a narrowing of the $n(x,y)$ since the dipoles attract each other in the $x$-$y$ plane assembling in a side-by-side configuration. 
Notice that the width of $n(x,y)$ at $\phi>\phi_m$, e.g. $\phi=90^{\circ}$ in Fig.~\ref{fig_quasi_2d_den2}$(\text{a}_3)$, is still larger than the one for $\phi=0$ [Fig.~\ref{fig_quasi_2d_den2}$(\text{a}_1)$]. 
This occurs because the increase of the width of the $n(x,y)$ until $\phi=\phi_m$ is larger than its shrinking for $\phi>\phi_m$. 
This is in accord with the energy minimization in Eq.~(\ref{avgddi}). 
Moreover, we observe that $n(y,z)$ changes from being highly elongated along $z$ [Fig.~\ref{fig_quasi_2d_den2}($\text{a}_4$), ($\text{a}_5$)] if $\phi<\phi_m$, to being so across $y$ [Fig.~\ref{fig_quasi_2d_den2}($\text{a}_6$)] for $\phi>\phi_m$ due to the sign change of the DDI along the $z$-axis.

\subsection{Shaping the droplet lattice}

\begin{figure}
\centering
\includegraphics[width = 0.48\textwidth]{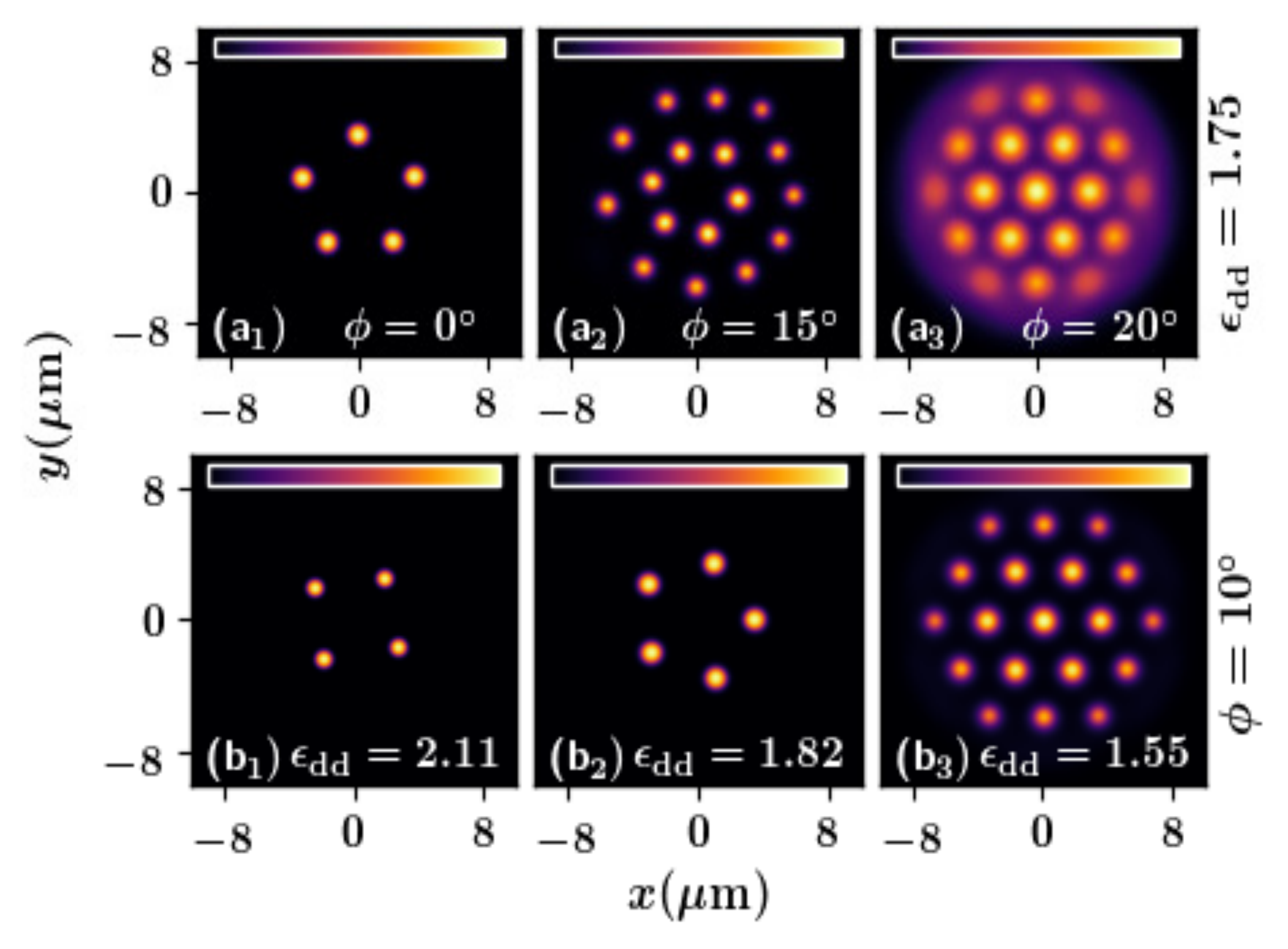}
\caption{Integrated density profiles $n(x,y)$ of the dBEC containing $N = 2.5  \times 10^{5}$ atoms with $\epsilon_{dd}=1.75$ for (a$_1$) $\phi = 0^{\circ}$, (a$_2$) $\phi = 15^{\circ}$, and (a$_3$) $\phi = 20^{\circ}$. $n(x,y)$ for (b$_1$) $\epsilon_{dd}=2.11$, (b$_2$) $\epsilon_{dd}=1.82$ and (b$_3$) $\epsilon_{dd}=1.55$ with a fixed $\phi = 10^{\circ}$. 
Various bound state configurations can be created by either tuning the tilt angle, e.g. hexagonal lattices, or adjusting $\epsilon_{dd}$ such as different polygons. 
The external quasi-2D geometry is characterized by $(\omega_x, \omega_y, \omega_z) = 2 \pi \times (45, 45, 133) \rm Hz $ defining a length scale $l_{{\rm osc}}=1.17 \mu m$. The colorbar (units of $1/l_{{\rm osc}}^2=0.73 \mu m^{-2}$) refers to the density which is distinct for each panel and has a gradient from zero (black) to a maximum value (yellow).}
\label{fig_phi_den}
\end{figure}

In a similar vein, the anisotropy of the dipole-dipole interaction substantially impacts also the $\rm DL_{M}$ and $\rm SS$ phases, which emerge {\it only} when the dipoles are repulsive (in the $x$-$y$ plane), i. e. $\phi < \phi_m$. 
Below, we focus on the deformations of the $\rm DL_{M}$ phase from variation of either $\phi$ or $\epsilon_{dd}$ through $a_s$. In Fig.~\ref{fig_phi_den} (a$_1$)-(a$_3$), $\epsilon_{dd}=1.75$ and $N = 2.5 \times 10^{5}$. The number of individual droplets in the lattice increases as $\phi$ is increased and the dipoles become less repulsive and weaker in magnitude across the $x$-$y$ plane, a process favoring further fragmentation of the droplets. 
Around $\phi =20^{\circ}$, a phase transition occurs with the emergence of a hexagonal SS phase. The SS state forms due to the effective weakening of the DDI for larger $\phi<\phi_m$ as compared to the contact interaction, see also the respective alterations in the phase diagram of Fig.~\ref{phae_diag_pan}](a), (b). 
The location of this phase boundary with respect to $\phi$ depends on the strength of the contact interaction.

Such a distribution has been independently confirmed by tuning of $a_s$ with a fixed magnetic field~\cite{adhikari_2022}. 
It is worthwhile to mention that the structural deformation of the droplet lattice can be  achieved independent of $a_s$ by changing the trap aspect ratio in the crossover from 2D to 1D, as  demonstrated in Ref.~\cite{poli2021}. 
Along these lines, by adjusting $\epsilon_{dd}$ (and in particular $a_s$) it is possible to create a variety of intriguing droplet patterns for fixed $\phi$ such as squares at $\epsilon_{dd}=2.11$  [Fig.~\ref{fig_phi_den}(b$_1$)], pentagons at $\epsilon_{dd}=1.82$ 
[Fig.~\ref{fig_phi_den}(b$_2$)] or hexagonal-type lattices at $\epsilon_{dd}=1.55$ [Fig.~\ref{fig_phi_den}(b$_3$)]. 
Recall that a decreasing $\epsilon_{dd}$ (with $\epsilon_{dd}>1.4$) favors the breaking of each individual droplet into multiple segments, see also 
Fig.~\ref{phae_diag_pan}($\text{a}$), since the DDI dominates with respect to the contact contribution. 
Exploring the energetics of such configurations, in analogy to what has been done, e.g., for multi-vortex configurations (see, e.g.,~\cite{zamp} for an example) would be a particularly intriguing direction for future study. 

For completeness, we remark that the background density of a SS becomes gradually denser for larger $\phi$, destroying its SS nature\footnote{The transition from the SS to the SF state can be equally seen in both the density and the momentum distribution of the dBEC. Indeed, periodic density undulations vanish as the SF is entered. 
Also, the momentum distribution of a SF is characterized by a single peak structure while for a SS multiple additional peaks appear.} and finally establishing a SF state\footnote{A corresponding SF state is essentially insensitive to variations of $\phi$ and only its width increases (decreases) for larger (smaller) $\phi$ as long as $\phi<\phi_m$ ($\phi>\phi_m$).}. 
Notice that the transition boundary from a SF to a SS phase can also be determined from the so-called contrast, $\mathcal{C} = (n_{\rm max} - n_{\rm min})/(n_{\rm max} + n_{\rm min})$. 
Here, $n_{\rm max}$ and $n_{\rm min}$ are the neighbouring density maxima and minima, respectively~\cite{T_Bland2022}. 
A SF state occurs for $\mathcal{C} = 0$, while $\mathcal{C} \ne 0$ corresponds to a density modulated state.

\section{Quench dynamics} \label{Quench_sub}

We next investigate the non-equilibrium dynamics by initializing  the dBEC in a SF state (with $\epsilon_{dd}=1.1$) and following  
an interaction quench to larger $\epsilon_{dd}$ values such that the SS or the droplet phase is dynamically entered. 
The spontaneous nucleation and properties of these beyond mean-field structures are studied in quasi-2D and quasi-1D geometries~\cite{Fabian_2019, Ferrier_2016} for $\phi=0^{\circ}$. Note that even for $\phi \neq 0$ the dynamics is not substantially altered.

\subsection{Dynamical nucleation of 2D SS and DL lattices}\label{Quench2D_dyn_sub}

\begin{figure}
\includegraphics[width=0.48\textwidth]{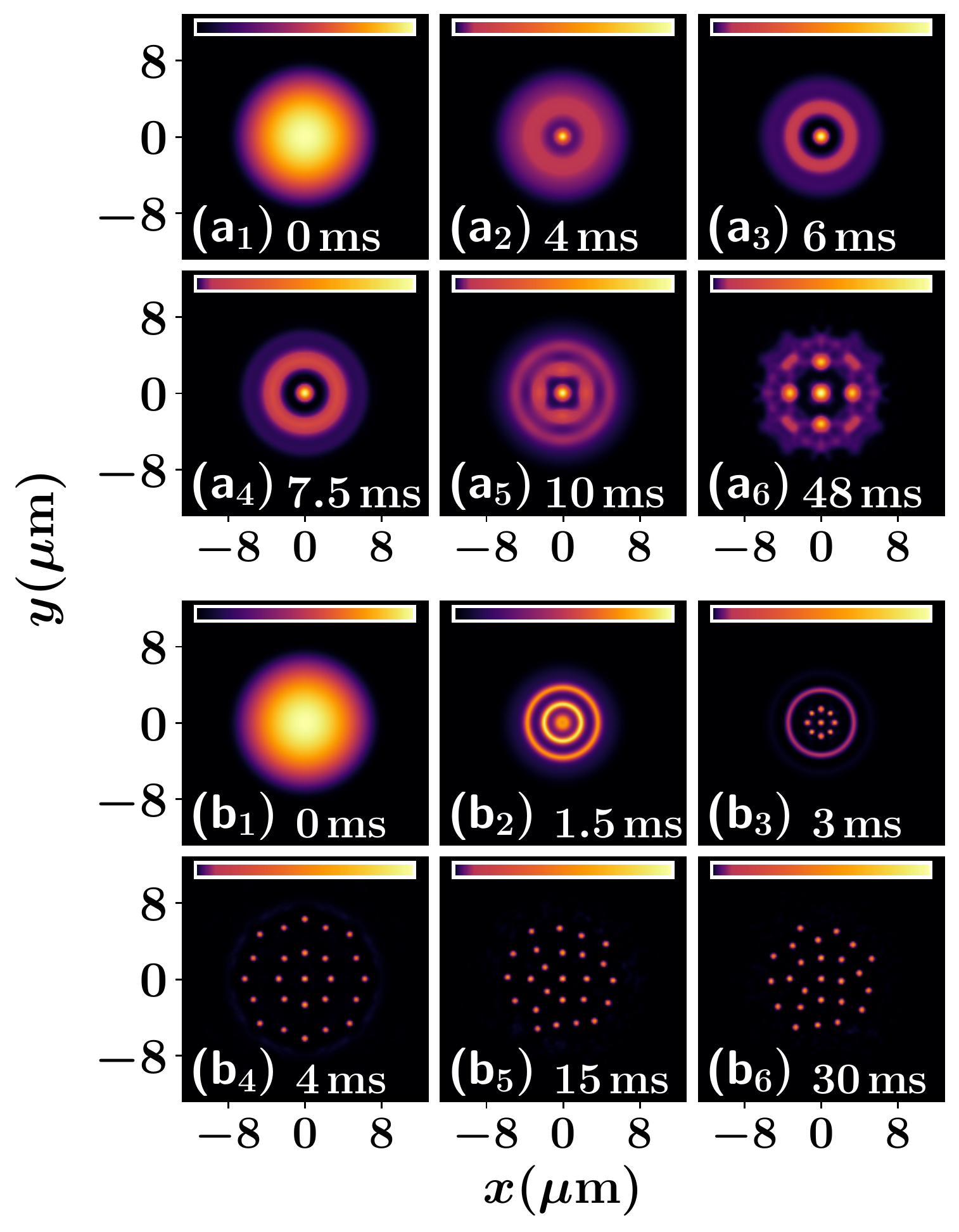}
\caption{Snapshots of the dBEC density $n(x,y)$ following an interaction quench from a SF state with $\epsilon_{dd}=1.1$ to ($\text{a}_1$)-($\text{a}_6$) $\epsilon_{dd}=1.45$ and ($\text{b}_1$)-($\text{b}_6$) $\epsilon_{dd}=1.87$. 
The initially 2D smooth density profile deforms in the course of the evolution towards a SS and a droplet lattice respectively. 
The dBEC consists of $N=6 \times 10^{4}$ atoms and it is confined in a quasi-2D harmonic trap characterized by $(\omega_x, \omega_y, \omega_z) = 2\pi \times (45, 45, 133) \rm Hz$. 
The colorbar denotes the density in units of $1/l_{{\rm osc}}^2=0.73 \mu m^{-2}$ and differs for each panel while featuring a gradient from zero (black) to a maximum value (yellow). 
The system's characteristic timescale set by the trap is $\omega_x^{-1}=3.5 \rm ms$ and the length scale defined through the harmonic oscillator length $l_{{\rm osc}}=1.17 \mu m$.}
\label{den_pan_quench}
\end{figure}

Representative instantaneous density profiles $n(x,y;t)$ of the quasi-2D dBEC are presented in Fig.~\ref{den_pan_quench}($\text{a}_2$)-($\text{a}_6$) after a quench from a SF state with $\epsilon_{dd}=1.1$ to a SS having  $\epsilon_{dd}=1.45$ according to the phase diagram of Fig.~\ref{phae_diag_pan}(a). 
The initially smooth 2D TF distribution $n(x,y;t=0)$  [Fig.~\ref{den_pan_quench}($\text{a}_1$)]   
is dynamically modified due to the quench. 
Indeed, the roton~\cite{Hertkorn_2021_supersolidity,Jona-Lasinio_2013}
induced softening in the post-quench phase seeds the
subsequent pattern formation. 
As a result, ring-shaped density structures develop 
becoming more pronounced as time-evolves [Figs.~\ref{den_pan_quench} ($\text{a}_2$)-($\text{a}_5$)]. 
For an analysis of the roton-induced dynamics and its shape in a 3D harmonically trapped dBEC see Ref.~\cite{Schmidt_2021}. 
It is, in fact, the progressive growth of the roton mode, characterized by a non-zero angular momentum, which is responsible in the early time dynamics ($t\sim \omega_x^{-1}$) for the development of these ring structures accompanied by density depleted regions. 
Later on, for $t > 20 {\rm ms} > \omega_x^{-1}$, following the interference of the ring densities (stemming from the radial roton) and the growth of azimuthal undulations (originating from the angular roton\footnote{The roton modes in our quasi-2D harmonically trapped setup are characterized by the quantum number $m$. In this sense, $m=0$ refers to the radial roton manifesting as a ring structure and $m \neq 0$ are the angular rotons having a 
corresponding number of azimuthal nodes~\cite{Schmidt_2021}.}) the dBEC distribution splits into four overlapping density peaks arranged in a  
square configuration and surrounding the central density hump which appears to be isolated from the others, see Fig.~\ref{den_pan_quench}($\text{a}_6$). 
At these timescales ($t \gg \omega_x^{-1}$), quantum fluctuations take over and the roton-induced growth saturates~\cite{Chomaz_2019, Fabian2019}.

\begin{figure}
\includegraphics[width=0.48\textwidth]{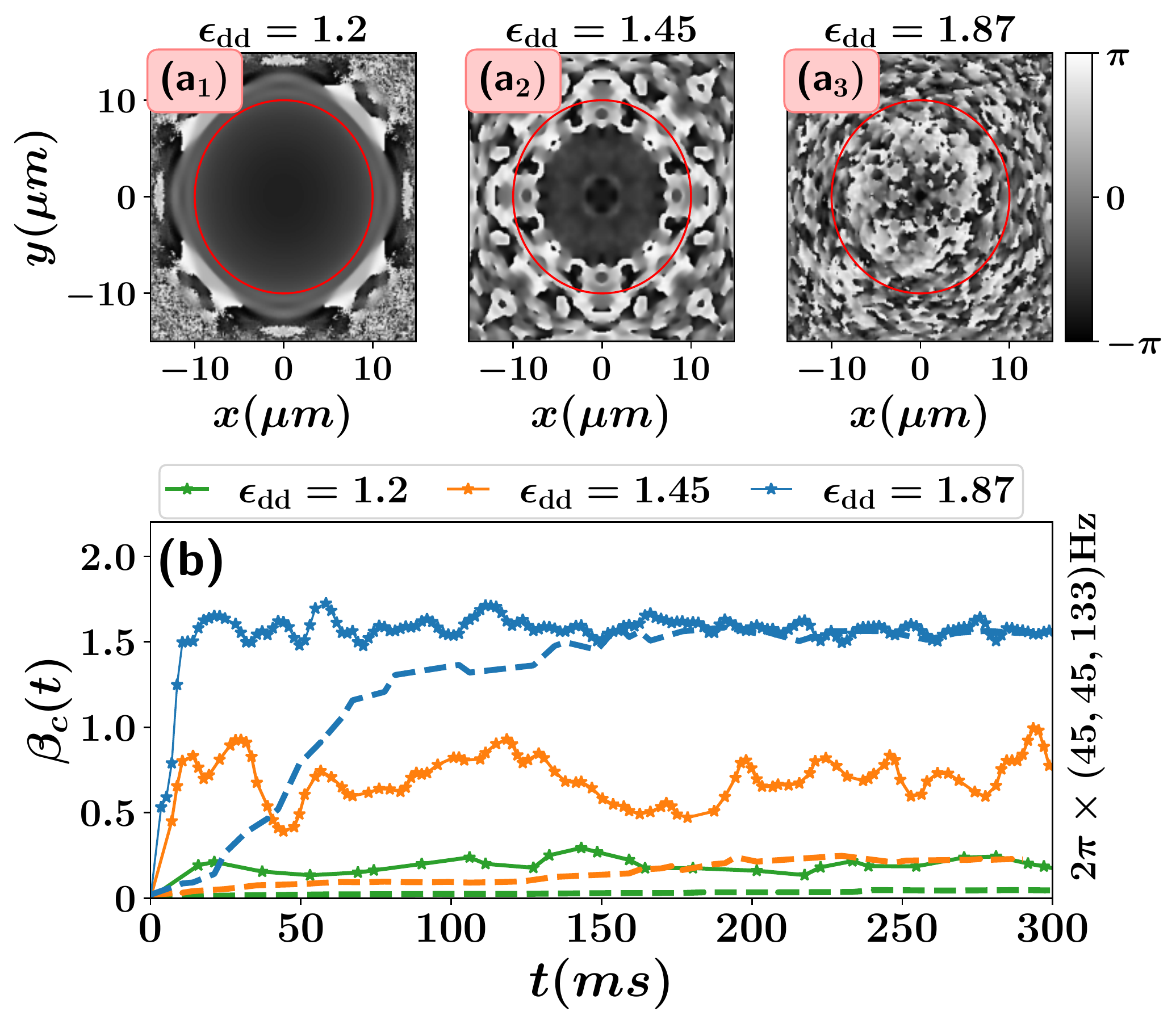}
\caption{Characteristic phase profiles ($\text{a}_1$)-($\text{a}_3$) $\Phi (x,y,z=0)$ at a specific time instant (see the legends) in the long-time dynamics after the quench. 
The pink circles designate the edges of the dBEC cloud. In all cases a quench of a quasi-2D dBEC from its SF state at $\epsilon_{dd}=1.1$ is considered towards the ($\text{a}_1$) SF with $\epsilon_{dd}=1.2$, ($\text{a}_2$) the SS having $\epsilon_{dd}=1.45$ and ($\text{a}_3$) the $\rm DL_{S}$ with $\epsilon_{dd}=1.87$ phase. 
The phase undulations designate the SS and the droplet states in contrast to the smooth phase of a SF. 
(b) The time-evolution of the global phase coherence $\beta_c(t)$ is presented for different post-quench $\epsilon_{dd}$ values in the quasi-2D geometry with $(\omega_x, \omega_y, \omega_z) = 2 \pi \times (45, 45,133)$Hz. 
The dashed lines in (b) represent the time-evolution of $\beta_c(t)$ when $\epsilon_{dd}(t)$ increases in a linear manner with ramp time $\tau=120 {\rm ms} \gg \omega_x^{-1}$. 
Coherence is completely lost in the droplet regime, while it is almost perfectly maintained following an adiabatic ramp towards the SF and the SS phases. The
colors of the dashed lines refer to the same post-quench $\epsilon_{dd}$ values with the ones indicated by the solid lines. 
Other system parameters are the same as in Fig.~\ref{den_pan_quench}. }
\label{global_coh}
\end{figure}

The phase profile $\Phi = \tan^{-1}\big[\rm Im(\psi(\vb{r)})/\rm Re(\psi(\vb{r}))\big]$, where $\rm Re(\psi(\vb{r}))$ [$\rm Im(\psi(\vb{r}))$] is the real [imaginary] part of the wave function, can be seen in $\Phi(x,y,z= 0)$ [Fig~\ref{global_coh} ($\text{a}_2$)] at $t= 150 {\rm ms} \gg \omega_x^{-1}$, see in particular the spatial region marked by the pink circle within which the SS resides. 
The enhanced distortions at the condensate edges ($\abs{x} \leq 8\mu m$, $\abs{y}\leq 8 \mu m$) stem from the highly non-smooth low density in conjunction with the high frequency breathing of the entire cloud. 
Actually, there are two breathing modes, see also Appendix~\ref{breathing} and Fig.~\ref{Quench_rms_pan}(a$_3$), (a$_4$). In particular, one of the breathing modes hardens in the SS state, similarly to the radial roton mode~\cite{Hertkorn_2021_supersolidity}, for increasing $\epsilon_{dd}$. 
This breathing mode is related to the spatially modulated SS density. The {frequency of the} second mode is linked to the background SF and diminishes gradually with increasing $\epsilon_{dd}$. 
The above behavior of $\Phi$ is to be contrasted with the normal SF exhibiting a uniform phase as demonstrated in Fig.~\ref{global_coh}($\text{a}_1$).

Next, we tune to a post-quench $\epsilon_{dd}=1.87$, where a $\rm DL_{S}$ phase forms. The dBEC is again dynamically distorted, already for $t<\omega_x^{-1}$, showing a two-ring structure and a central density hump due to the presence of the roton mode. 
The latter grows at a faster rate compared to the $\epsilon_{dd}=1.45$ quench, see Fig.~\ref{den_pan_quench}($\text{a}_2$) and ($\text{b}_2$), as expected by the underlying excitation spectrum~\cite{Hertkorn_2021_supersolidity}.  
This leads  to the disintegration of the inner ring and the central density peak into multiple droplets around $t \sim \omega_x^{-1}$ [Fig.~\ref{den_pan_quench}($\text{b}_3$)]. 
At the outer rim of the dBEC, a low density circular structure appears at large radii, emerging from the edges of the cloud [Fig.~\ref{den_pan_quench}($\text{b}_3$)]. 
This metastable configuration subsequently ($t>\omega_x^{-1}$) breaks into a droplet lattice\footnote{The position of the roton minimum $k_{\rm rot}$ should satisfy $k_{\rm rot} l_z \ge 1$, where $l_z = \sqrt{\hbar/ m \omega_z}$ is the harmonic oscillator length scale along the tightly confined direction. 
In our case, $k_{\rm rot} = 3.39 \mu m ^{-1}$ with $l_z = 0.68 \rm \mu m$. 
Similar findings have been reported e.g. in the experiment of Ref.~\cite{Chomaz_2018} where $k_{\rm rot} = 2.5 \mu m ^{-1}$ and $l_z = 0.625 \rm \mu m$.} [Fig.~\ref{den_pan_quench}($\text{b}_4$)-($\text{b}_6$)]. 
It should be emphasized that while the post-quench ground-state represents a $\rm DL_{S}$, we encounter here the spontaneous nucleation of a $\rm DL_{M}$. 
This droplet cluster features, at the early times  ($t \geq \omega_x^{-1}$), a global breathing motion and thus the distance between individual droplets changes [Fig.~\ref{den_pan_quench}($\text{b}_5$)-($\text{b}_6$)]. 
Let us note that due to the lack of background SF density, only one breathing mode connected to localized density arrays exists here. 
However, in the long-time dynamics (roughly $t> 70 {\rm ms} \gg \omega_x^{-1}$), the breathing amplitude reduces and the cluster remains practically stationary, see also Appendix~\ref{breathing}, at least up to $500$ ms, indicating that the system is in a prethermalized state.

The number of isolated droplets contained in the cluster in the long-time dynamics ($t \gg \omega_x^{-1}$) increases for larger post-quench $\epsilon_{dd}$\footnote{It is worth mentioning that the number of dynamically nucleated droplets in the long-time evolution is, in general, larger from the one of the respective ground-state configuration. For instance, in the case of $\epsilon_{dd}=1.75$ the droplet lattice contains twenty four individual droplets in contrast to four formed in the ground-state for $N = 10^5$.}, namely deeper in the droplet regime, see in particular Table~\ref{Table_1} for $\phi=0^{\circ}$. 
In contrast to this response their number realized for a fixed post-quench $\epsilon_{dd}$ is smaller upon increasing the 
tilt angle. 
This trend is visualized in the case examples of Table~\ref{Table_2} and it is attributed to the reduced magnitude of the associated DDI. 

We also note in passing that independently of the post-quench  $\epsilon_{dd}$ {regular} polygonal lattices do not form, in agreement with~\cite{Santos_2016}, but only crystalline structures as the one depicted in Fig.~\ref{den_pan_quench}($\text{b}_6)$. 
The existence of the droplet clusters can also be distinguished by inspecting their phase profiles which are highly non-uniform [Fig.~\ref{global_coh}($\text{a}_3$) at $t= 200 {\rm ms} \gg \omega_x^{-1}$] even when compared with the SS phase [Fig.~\ref{global_coh}($\text{a}_2$)]. 

\subsection{Control of the global phase coherence during the evolution}

An observable that provides further verification for the dynamical creation of the above-discussed beyond mean-field states is the so-called global phase coherence~\cite{Tanzi_2019,Ilizhfer2021,Chomaz_2019}, see also Fig.~\ref{Schematic}(c), which is defined as follows 
\begin{eqnarray}\label{GPC}
\beta_c(t) = \int d\vb r \big[n(\vb{r,t})\abs{(\Phi(\vb{r},t) - \Bar{\Phi}(t))}\big]. 
\end{eqnarray}
Here, $\Bar{\Phi}(t)$ is the spatially averaged phase. 
Accordingly, following adiabatic pulses a SF or a SS state has perfect global phase coherence ($\beta_c(t)=0$), while self-bound quantum droplets feature $\beta_c (t) \ne 0$. 

This is expected since as we previously argued the phase of the gas becomes highly distorted for droplet states in contrast of being almost uniform for SS and SF phases. 
The dynamics of $\beta_c (t)$ is provided in Fig.~\ref{global_coh}($\text{b}$) for various post-quench values of $\epsilon_{dd}$. 
The pre-quench (initial) SF state is perfectly coherent, i.e. $\beta_c(t=0)=0$. 
In all cases, the increase of $\beta_c (t)$ at short timescales ($t \sim \omega_x^{-1}$) is inherently related to the quench protocol and becomes more enhanced for larger quench amplitudes where the respective import of energy into the system is naturally larger. 

For quenches within the SF phase, e.g., $\epsilon_{dd}=1.2$, $\beta_c(t)$ fluctuates, within $0.13< \beta_c(t) < 0.2$ after $t>0.7 {\rm ms}$, due to the breathing motion of the gas originating from the quench, see also Appendix~\ref{breathing} and Fig.~\ref{Quench_rms_pan}. 
For the same reason, $\beta_c(t>0)$ is finite also for quenches towards the SS regime ($\epsilon_{dd}=1.45$) while the relatively intensified oscillations compared to the SF stem from the arguably larger amplitude of the underlying breathing mode. 
The coherence loss in the SS phase can be mitigated by adiabatically increasing $\epsilon_{dd}$\footnote{Note that a non-adiabatic ramp of $\epsilon_{dd}(t)$, e.g., in our case realized for $\tau < 150 \ \rm ms$, always results in a finite global phase coherence.}. In support of this argument, we utilize a linear ramp $\epsilon_{dd}(t)= (\epsilon^{F}_{dd} - \epsilon^{I}_{dd})t/\tau$, with $\epsilon^F_{dd}$ ($\epsilon^I_{dd}$) being the final (initial) relative strength, while $\tau$ is the ramp time. 
An almost adiabatic increase of $\epsilon_{dd}(t)$ is achieved for $\tau=120 {\rm ms} \gg \omega_x^{-1}$ with $\beta_c(t) \to 0$ for the SF and SS as shown in Fig.~\ref{global_coh}($\text{b}$). 
Turning to quenches in the droplet regime, for instance in the case of $\epsilon_{dd}=1.87$, we observe that $\beta_c(t)$ features an increase at short timescales and then fluctuates around $\pi/2$. 
This saturation tendency of $\beta_c(t)$ deep in the evolution holds equally when performing a linear increase of $\epsilon_{dd}(t)$, see Fig.~\ref{global_coh} ($\text{b}$). 
Note that $\beta_c(t)$ can be at most $\pi/2$ as has also been demonstrated in Ref.~\cite{Tanzi_2019}.  
This response implies that the initial phase coherence of the SF phase is rapidly lost and the individual droplets become highly incoherent.

\subsection{Generation of droplet and SS arrays in the quasi-1D regime}\label{Quasi1D_dyn_sub} 

\begin{figure}
\includegraphics[width=0.48\textwidth]{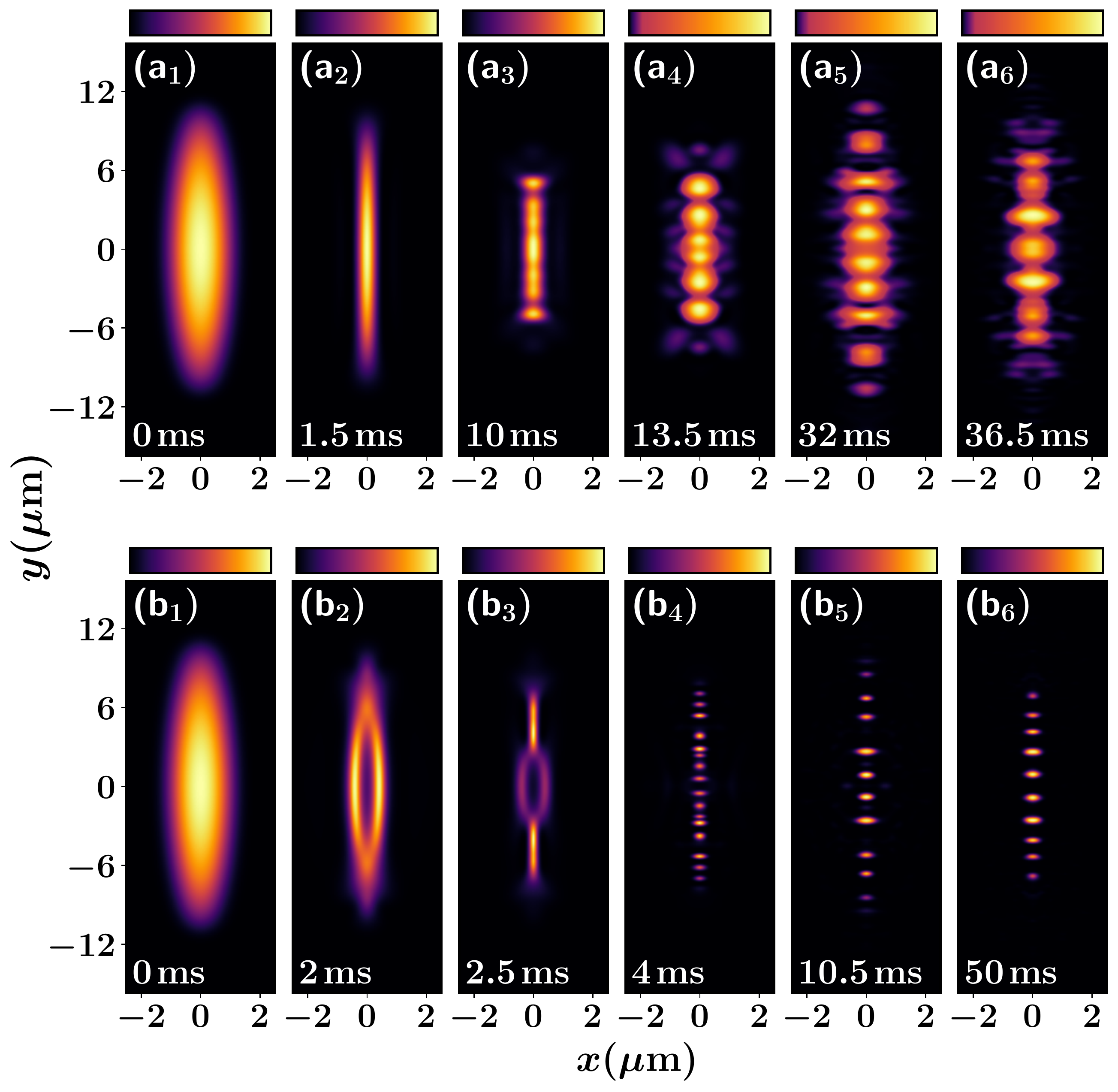}
\caption{The same as in Fig.~\ref{den_pan_quench} but for the quasi-1D setting. 
The generation of ($\text{a}_1$)-($\text{a}_6$) SS for $\epsilon_{dd}=1.45$ and ($\text{b}_1$)-($\text{b}_6$) droplet arrays when $\epsilon_{dd}=1.87$ is illustrated. 
The dBEC consists of $N  =6 \times 10^{4}$ atoms and it resides in a quasi-1D trap with $(\omega_x,\omega_y,\omega_z) = 2\pi \times (227,37, 135) \rm Hz$. 
The colorbar in units of $1/l_{{\rm osc}}^2= 1.33 \mu m^{-2}$ refers to the density while it is different for each panel and changes from zero (black) to a maximum value (yellow). The characteristic length scale is the harmonic oscillator length  $l_{{\rm osc}}=0.85 \mu m$ and the corresponding timescale is $\omega_x^{-1}=0.7 \rm ms$.}
\label{den_axial_quench}
\end{figure}

Subsequently, we study the quench-induced dynamics of the axially elongated dBEC. 
Characteristic density snapshots across the $x$-$y$ plane when performing a quench within the SS phase, e.g. $\epsilon_{dd}=1.45$, are presented in 
Fig.~\ref{den_axial_quench}($\text{a}_1$)-($\text{a}_6$). 
The quasi-1D TF distribution [Fig.~\ref{den_axial_quench}($\text{a}_1$)] of the initial SF state (with $\epsilon_{dd}=1.1$) experiences prominent spatial deformations due to the ensuing roton dynamics~\cite{Hertkorn_2021_supersolidity} being activated when crossing the SF to SS phase boundary. 
Specifically, the increased post-quench $\epsilon_{dd}$ enforces a contraction [Fig.~\ref{den_axial_quench}($\text{a}_2$), ($\text{a}_3$)] and expansion [Fig.~\ref{den_axial_quench}($\text{a}_4$)-($\text{a}_6$)] of the entire cloud along both $x$ and $y$ directions. 
This collective motion prevails at short timescales [Fig.~\ref{den_axial_quench}($\text{a}_2$)] but afterwards 
spatial modulations in the density profile arise along the $y$-axis [Fig.~\ref{den_axial_quench}($\text{a}_3$)-($\text{a}_6$)]. 
For instance, six density peaks are detected at $t={\rm 13.5 ms} >\omega_x^{-1}$ [Fig.~\ref{den_axial_quench}($\text{a}_4$)] which break into several ones at $t=32 {\rm ms} \gg \omega_x^{-1}$ [Fig.~\ref{den_axial_quench}($\text{a}_5$)]. 
These arrays of overlapping density humps developing in $n(x,y)$ reveal the dynamical formation of the SS state. 
It is also worth mentioning that the periodic spatial compression and expansion of the dBEC is characterized similarly to the quasi-2D case by two distinct frequencies. 
We remark that the participation of the two distinct breathing frequencies is a characteristic of the emergence of the SS phase, a result which has been evinced independently in Ref.~\cite{Tanzi_2019}. 
 
Utilizing a quench to $\epsilon_{dd}=1.87$ leads to a dramatically different response of the dBEC as shown in Fig.~\ref{den_axial_quench}($\text{b}_1)$-$(\text{b}_{6})$. 
Already at the early stages of the evolution $\sim \omega_x^{-1}$ we observe that the original smooth density configuration [Fig.~\ref{den_axial_quench}($\text{b}_1$)] transforms into an elliptic halo 
profile [Fig.~\ref{den_axial_quench}($\text{b}_2$)]. The width of the latter progressively shrinks across the transverse $x$-direction [Fig.~\ref{den_axial_quench}($\text{b}_3$)] until the entire cloud becomes highly elongated breaking into an array of droplets [Fig.~\ref{den_axial_quench}($\text{b}_4$)]. 
The reason behind the formation of the elliptic halo 
states is the arising modulational instability due to admixture of different roton modes discussed, for instance, in Refs.~\cite{Hertkorn_2021_supersolidity, Hertkorn_2021_density_fluc, Schmidt_2021} and triggered herein by the quench within the SS phase. 
As a by-product, the dBEC fragments into multiple highly localized peaks organized in a crystal pattern. 
\begin{table}
\large
\centering
\begin{tabular}{l | a | b | a | b}
\hline
\rowcolor{LightCyan}
\mc{1}{Trap geometry}  \vline& \mc{1}{$\epsilon_{dd}=2.18$} \vline& \mc{1}{$2.01$} \vline& \mc{1}{$1.87$} \vline& \mc{1}{$1.75$} \\
\hline
quasi-2D & 27 & 25 & 24 & 17 \\\hline
quasi-1D & 16 & 15 & 13 & 11 \\ \hline
\end{tabular}
\caption{Number of isolated droplets in the quasi-2D and the quasi-1D regimes for different post-quench $\epsilon_{dd}$ values and fixed $\phi = 0^{\circ}$. 
The number of droplets contained in a cluster becomes larger for increasing $\epsilon_{dd}$, i.e., deeper in the droplet phase. Initially, the dBEC resides in a SF state characterized by $\epsilon_{dd}=1.1$.}\label{Table_1}
\end{table}

Notice that in contrast to the SS case of post-quench $\epsilon_{dd}=1.45$, these density humps are entirely isolated and comprise the self-bound droplets. 
The droplet array becomes stationary in the long time dynamics e.g. $t> 100 {\rm ms} \gg \omega_x^{-1}$, while at earlier times ($t>\omega_x^{-1}$) the inter-droplet distance changes, see e.g. Fig.~\ref{den_axial_quench}($\text{b}_4$)-($\text{b}_6$). 
This phenomenon can be traced back to the collective breathing motion of the cloud due to the interaction quench. 
Interestingly, unlike the SS phase, here a single breathing frequency occurs, a property that is attributed to the crystal nature of the droplet phase. 
These properties of the breathing mode of a SS and a droplet have also been experimentally observed for an elongated dBEC in Ref.~\cite{Tanzi2019}. 
It is also worth mentioning that $\beta_c (t)$ (not shown) features a response similar to the one of the quasi-2D setup. 
Namely, in the droplet regime it is eventually ($t \gg  \omega_x^{-1}$) maximally lost, while for the SF and SS phases it acquires relatively smaller finite values (due to the quench) while being minimized following an adiabatic ramp of $\epsilon_{dd}$. 
We finally note that as in the quasi-2D scenario, the stationary array has a lesser number of droplets upon reducing post-quench $\epsilon_{dd}$ or increasing $\phi$ as shown in Table~\ref{Table_1} and Table~\ref{Table_2}.

\begin{table}[h]
\large
\centering
\begin{tabular}{l | a | b | a | b | a|}
\hline
\rowcolor{LightCyan}
\mc{1}{Trap }  \vline& \mc{1}{$\phi=0^{\circ}$} \vline& \mc{1}{$5^{\circ}$} \vline& \mc{1}{$10^{\circ}$} \vline& \mc{1}{$15^{\circ}$} \vline& \mc{1}{$20^{\circ}$} \\
\hline
Quasi-2D & 27 & 26 & 25 & 24 & 17 \\ \hline
Quasi-1D & 16 & 15 & 14 & 10 & 8  \\ \hline
\end{tabular}
\caption{Number of isolated droplets in the quasi-2D and the quasi-1D regimes for a specific post-quench $\epsilon_{dd}=2.18$ and considering different field orientations $\phi^{\circ}$. 
In both cases the amount of individual droplets in the respective lattice decreases as $\phi$ increases due to the effective weakening of the DDI. Initially, the dBEC resides in a SF state characterized by $\epsilon_{dd}=1.1$.}\label{Table_2}
\end{table}

\section{Three-body loss and self-evaporation of SS and droplet phases}\label{TBI_append}

A central obstacle for the detection of self-bound structures, characterized by highly localized densities, is that experimentally~\cite{Chomaz_2016, Fabian_2019} they suffer from three-body losses. 
In the following, we explain the impact of the underlying three-body loss rate in the formation of SS and droplet quasi-2D configurations. 
Contrary to previous studies, the employed rotating magnetic field [Eq.~(\ref{avgddi})] allows us to expose the effect of the  underlying losses in different interaction regimes compared to the dipolar interaction length. 
The corresponding eGPE~\cite{chomaz2022dipolar} has the form of Eq.~(\ref{eGPE}) with the additional imaginary contribution $- (i \hbar K_3 /2) \abs{\psi(\textbf{r},t)}^4 \psi(\vb{r},t)$, where $K_3$ denotes the three-body recombination rate~\cite{chomaz2022dipolar}. 
A detailed discussion regarding the competition between the three-body recombination and beyond mean-field processes is provided in Ref.~\cite{chomaz2022dipolar}.

The important point here is that the scaling of the three-body recombination rate is obtained in terms of  $D=3a_{dd}(3\cos^2 \phi-1)/4$, where $a_{dd}=131 \ a_B$ herein. 
It was shown~\cite{ticknor2010three} that $K_3 \sim  D^4$ for $\epsilon_{dd} \gg a_{dd}/D$ ($a_s \ll D$) and $K_3 \sim \mathcal{C} a_s^2(a_s^2+\beta D^2)$ for $\epsilon_{dd} \ll a_{dd}/D$ ($a_{s} \gg D$). 
In the above expressions, the constants $\mathcal{C}=3! 32\sqrt{3} \pi^2 \hbar/m$ and $\beta \approx 0.44$. 
It should be emphasized that for $\phi=0^{\circ}$, both the SS and the droplet phases occur within $\epsilon_{dd} \gg a_{dd}/D$. 
This is exactly the situation  that has been considered thus far in the literature for interpreting experimental data~\cite{Chomaz_2018, Ferrier_2016}. 
Here, this is realized by  $K_3=7.1\times10^{-43} m^6/s$. 
However, since our dBEC is subjected to a tilted magnetic field, it is also possible to enter the regime $\epsilon_{dd} < a_{dd}/D$, where $K_3$ explicitly depends on $a_s$.

Accordingly, below, we will discern among these two important scenarios. 
Utilizing a post-quench $\epsilon_{dd}$ lying in the droplet regime for $\phi=0^{\circ}$, we adjust $\phi$ towards $\phi_m$ and a SF state forms, since the dipolar interaction is not strong enough to create droplets. Then, $K_3$ becomes $a_s$-dependent. 
On the other hand, when $\phi  \ll \phi_m$, and $\epsilon_{dd} > a_{dd}/D$, the loss-rate scales explicitly with $D$ and hence $\phi$. 
This tunability provides an additional knob for controlling the lifetime of the dynamically accessed self-bound states. 

\begin{figure}
\includegraphics[width=0.49\textwidth]{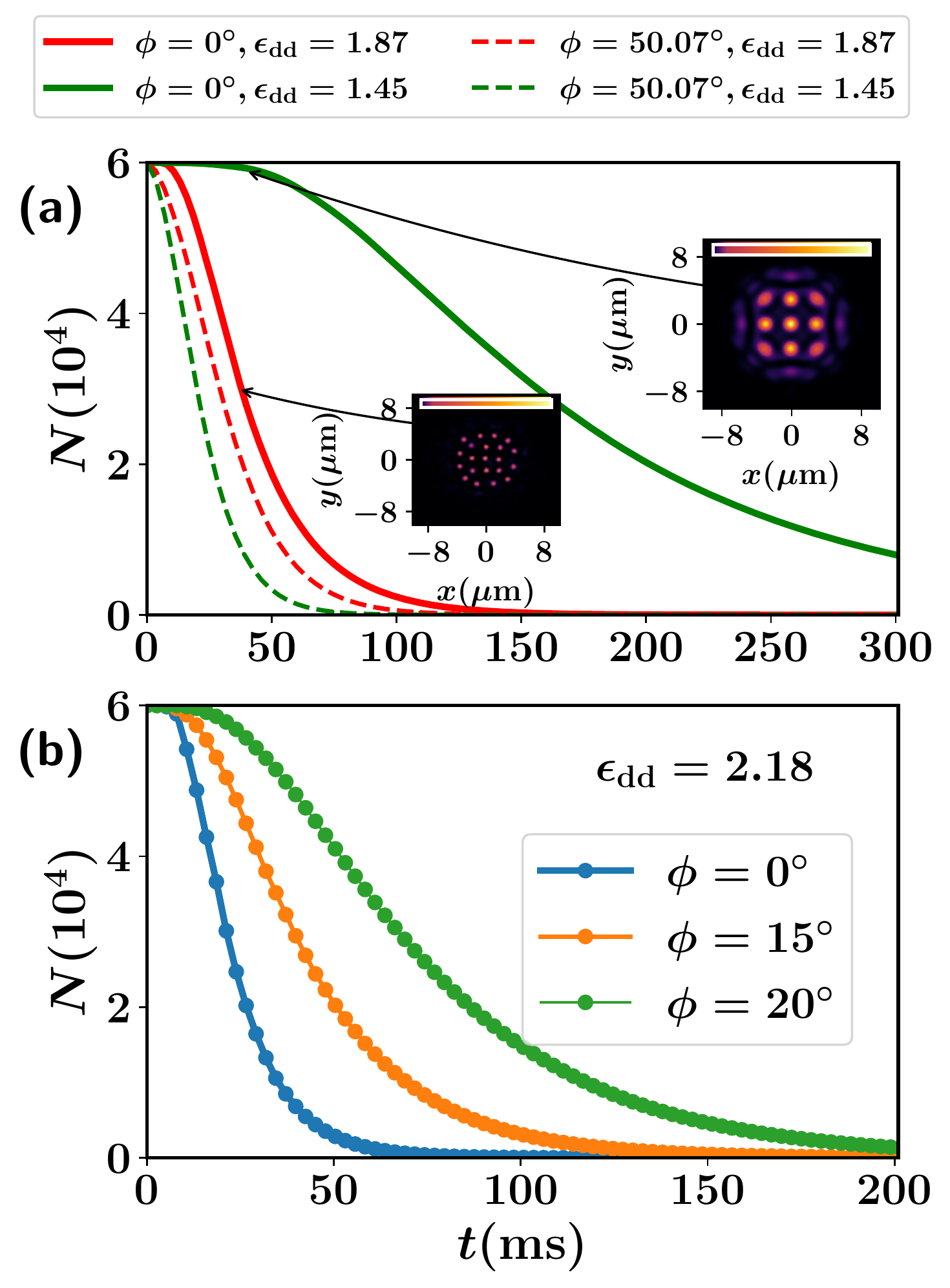}
\caption{(a) Dynamics of the particle loss for different post-quench values of $\epsilon_{dd}$ and tilt angles $\phi$ (see legend). 
Droplets feature the largest losses at $\phi=0$, while a finite field orientation leads to a SF state exhibiting more dramatic decay processes. 
Insets: Corresponding density snapshots $n(x,y)$ after the quench from $\epsilon_{dd}=1.1$ visualizing the formation of SS (upper panel) and droplets (lower panel) at short timescales. 
(b) Same as in (a) but for several $\phi$ (see legend) and a fixed post-quench $\epsilon_{dd}=2.18$ residing in the droplet phase. 
The emergent patterns persist for longer timescales by means of a tilted magnetic field. 
The droplet lifetime is enhanced for fixed $\epsilon_{dd} < a_{dd}/D$ and increasing $\phi$ as long as $\phi< \phi_m$. 
The dBEC with $N=6 \times 10^{4}$ experiences a quasi-2D trap of $(\omega_x, \omega_y,\omega_z)= 2\pi \times (45, 45, 133) \rm Hz$ setting a characteristic timescale $\omega_x^{-1}=3.5 \rm ms$.}
\label{TBI_pan_num}
\end{figure}

The atom losses are depicted in Fig.~\ref{TBI_pan_num} for various field orientations determined by $\phi$ and post-quench $\epsilon_{dd}$ leading to SS and droplet generation when $K_3=0$, see also Fig.~\ref{den_pan_quench}($\text{a}_1$)-($\text{a}_6$) and Fig.~\ref{den_pan_quench}($\text{b}_1$)-($\text{b}_6$). 
Throughout these cases, the dBEC is initiated in a SF state with $\epsilon_{dd}=1.1$. 
The case of $\phi = 0^{\circ}$ corresponds to $\epsilon_{dd} > a_{dd}/D$ for both the SS and droplet regimes and therefore $K_3$ is independent of $a_s$. 
We observe the nucleation of SS structures and droplet lattices for $\epsilon_{dd}=1.45$ [inset of Fig.~\ref{TBI_pan_num}(a) upper panel] and $\epsilon_{dd}=1.87$ [inset of Fig.~\ref{TBI_pan_num}(a) lower panel] respectively at intermediate timescales, $t \geq \omega_x^{-1}$. 
These are, however, only transient configurations due to $K_3\neq 0$ associated with non-negligible atom losses.  
As such they subsequently coalesce to narrow density peaks around the trap center (not shown) and afterwards decay for longer evolution times ($t \gg \omega_x^{-1}$) since three-body losses compete with the LHY contribution\footnote{We note that a similar phenomenology occurs also when considering the value of $K_3=1.2\times10^{-41} m^6/s$ used in Ref.~\cite{Santos_2016}}. 
The enhanced atom losses occurring in the droplet phase as compared to the SS one manifest due to the relatively higher localized densities of the former; compare, in particular, the solid red and green lines in Fig.~\ref{TBI_pan_num}(a). 
Notice that similar timescales of atom losses, exploiting a fixed magnetic field, have been observed in the experiment~\cite{Chomaz_2019, Fabian2019}. 

Importantly, tuning the dipolar anisotropy allows us to enter the $\epsilon_{dd}< a_{dd}/D$ region where the loss rate is interaction dependent. 
As a paradigmatic case, herein, we consider $\phi \approx 50^{\circ}$ where the loss coefficients  
are $K_3 = 2.47 \times 10^{-40} \rm m^6/s $ and 
$K_3 = 6.67 \times 10^{-40} \rm m^6/s $ for 
$\epsilon_{dd}=1.87$ and 
$\epsilon_{dd}=1.45$, respectively. 
Notice that for $\phi \approx 50^{\circ}$ the dipolar interaction is not strong enough, thus leading to a SF state for $\epsilon_{dd}=1.87$ and $\epsilon_{dd}=1.45$.  
However, since $K_3$ is interaction dependent it possesses a greater value for smaller $\epsilon_{dd}$ (larger $a_s$) resulting in an amplified lossy process, as can be seen by comparing the dashed lines to the solid ones in Fig.~\ref{TBI_pan_num}(a). 
Moreover, atom losses are accelerated in the $\epsilon_{dd} < a_{dd}/D$ region due to the prevalent $K_3 \sim a_s^4$, contrast the dashed with the solid lines in Fig.~\ref{TBI_pan_num}(a). 

Remarkably, we find that the droplet lifetime can be prolonged as long as $\phi < \phi_m$ and $\epsilon_{dd} > a_{dd}/D$ by increasing $\phi$ [Fig~\ref{TBI_pan_num}]($b$)] while always residing in the droplet regime. 
As a characteristic example we consider a quench towards $\epsilon_{dd} = 2.18$, where $K_3 = 7.1 \times 10^{-43} m^6/s$, $K_3 = 4.66 \times 10^{-43} m^6/s$ and $K_3 = 3.28 \times 10^{-43} m^6/s $ for $\phi=0^{\circ}$, $\phi=15^{\circ}$ and $\phi=20^{\circ}$ respectively. 
We deduce that the lifetime of the droplet structures realized at $\epsilon_{dd}= 2.18$ is prolonged by increasing $\phi$ lying in the interval $\phi< \phi_m$.

\section{Conclusions $\&$ future perspectives}\label{conclusion}

In the present work
we examined the ground-state phase diagram and the non-equilibrium dynamics of a harmonically trapped dBEC. 
We exploited the inherent anisotropy of the dipolar interaction by applying a fastly rotating magnetic field. 
Out considerations were based on an extended GP equation (3D eGPE), including quantum fluctuations to leading order, and allowed
us to study the emergent phases from the delicate interplay of isotropic short-range and anisotropic long-range forces in the quasi-1D and quasi-2D trapped geometries. 

We found that for $\phi<\phi_m$, four different phases emerge as a function of the strength of the contact interaction and atom number. These include the $\rm SF$ typically occurring for $\epsilon < 1.4$ and $\mu>0$, the $\rm SS$ residing in the vicinity of $\mu=0$, as well as the $\rm DL_{S}$ and the $\rm DL_{M}$ characterized by $\mu<0$. 
A SF state exhibits a smooth density distribution in sharp contrast to SS and droplets where substantially modulated patterns emerge. 
These structures have spatial overlap in the SS phase and exhibit crystalline behavior deep in the droplet regime. 
The crystal arrangements correspond to droplet clusters in quasi-2D forming canonical polygons, an outcome that holds equally for $\phi>\phi_m$.  
The number of separated droplets in a  cluster increases by either increasing the contact interaction strength or for large atom numbers, and for decreasing $\phi$. 
Transitions among the above-described phases are achieved by appropriately tuning the $s$-wave scattering length or the atom number. 
For $\phi>\phi_{m}$, where the dipoles attract, the $\rm DL_{S}$ and SF phases solely form. 
Additionally, the $\rm DL_{S}$ phase features a broad 2D circular distribution in the $x$-$y$ plane when $\phi>\phi_{m}$.

The SS and droplet phases can also form dynamically upon a quench of the $s$-wave scattering length from an initial SF state. 
SS clusters and droplet lattices are identified in quasi-2D, whilst elongated arrays of SS and droplets form in quasi-1D. 
These states are nucleated due to the roton-induced
dynamics manifesting as ring-shaped excitations (in quasi-2D) or elliptic halos (in quasi-1D) at early evolution times and are accompanied by a collective breathing motion of the dBEC background caused by the quench.  
Soon after their formation, these structures deform into arrays or clusters. 

The number of droplets in a lattice is larger for smaller post quench contact interactions or a finite angle such that $\phi<\phi_m$. 
The SF is maximally coherent throughout evolution, while the droplet phase displays total loss of coherence. 
For the SS phase, the existence of a finite background SF leads to a
partial loss of coherence. 
Interestingly, we observe that following quenches to the $\rm DL_{S}$ phase the system relaxes to a lattice, i.e., the $\rm DL_{M}$ phase. 
Moreover, the number of individual droplets participating in a lattice arrangement is larger  as compared to the ground-state configuration.

We also considered the loss of these dynamical phases due to three-body recombination. While the usual loss rate coefficient scales with $K_3\propto D^4$, because we tune the dipole anisotropy, we are able to probe the phase dynamics also in a regime where $K_3\propto a_s^4$. 
We find that the loss rates are enhanced towards the droplet regime for fully repulsively aligned dipoles at $\phi=0$ due to density effects, leading eventually to their self-evaporation. 
Our results show that employing a tilted magnetic field where the loss coefficient depends on the dipolar length, it is possible to prolong the lifetime of droplets as long as $\phi<\phi_m$. Otherwise, the droplet region suffers faster lossy mechanisms than the other states irrespectively of whether the loss coefficient depends or not on the interaction. 

Motivated by observations~\cite{Tang_2015}, we have restricted our study to a particular driving regime, i.e. $\omega \ll \Omega \ll \omega_L$. However, it would be intriguing to explore the impact of a weak $\Omega \ll \omega$ rotating field, thus extending present findings to the case where the dipoles cannot instantaneously follow the external magnetic field. 
Importantly, a quantitative understanding of the pairwise interaction of
the droplets in this system and of the pattern formation on the basis of their effective interacting particle system~\cite{siambook}, would be particularly interesting and relevant in this context.
Furthermore, comparison of the results herein with effective
lower-dimensional equations describing quasi-1D or quasi-2D dBECs
could be of interest as well; see, e.g.,~\cite{PhysRevLett.99.140406} for a 1D example.

It should also be possible to investigate topological pattern formation in the ground-state phases utilizing a rotating frame of reference. 
Likewise, understanding the phase diagram of the dBEC in the presence of nonlinear excitations such as vortex complexes in quasi-2D or solitary waves in quasi-1D will be helpful. 
Furthermore, studying the impact of finite temperatures~\cite{mithun2021statistical,de2021thermal} in the dynamical nucleation of SS and droplet lattices is certainly an intriguing perspective. 
Here, the dependence of the LHY term on the temperature should be carefully considered. 
The quench dynamics of a mixture of dipolar condensates across the distinct phases, e.g., discussed in Refs.~\cite{bisset2021quantum,scheiermann2022catalyzation} 
is a more computationally demanding effort, yet one 
worthy of consideration.

\section*{Acknowledgements} 
K.M thanks Stephanie M. Reimann for insightful discussions.
S. I. M. and H.R.S. acknowledge support from the NSF through a grant for ITAMP at Harvard University. 
S. H., S.D., P.K.P and S.M. acknowledge MHRD, Govt. of India for the research fellowship. 
K.M. is financially supported by the Knut and Alice Wallenberg Foundation (KAW 2018.0217) and  the  Swedish  Research  Council. 
P. G. K. is supported by the National Science Foundation under Grant No. PHY-2110030 and under Grant No. DMS-2204702.
We acknowledge the “PARAM Shakti” (Indian Institute of Technology Kharagpur) --- a national supercomputing mission, Government of India for providing computational resources. 

\appendix

\section{Variational treatment}\label{VarCal_sub}
 
\begin{figure}
\begin{center}
\includegraphics[width=0.45\textwidth]{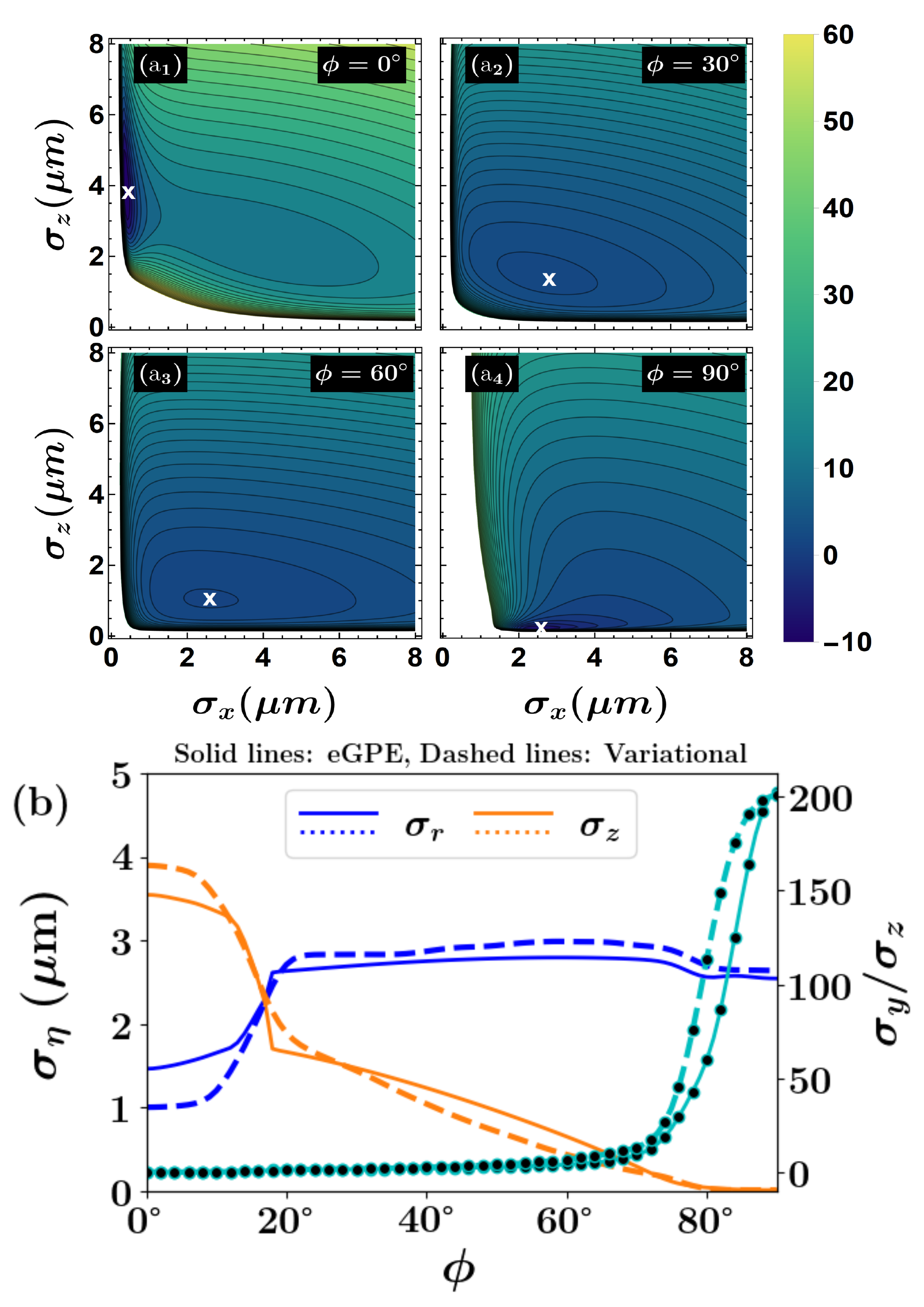}
\caption{Energy of the quasi-2D dBEC as predicted within the variational approach [Eq.~\eqref{energy_var}] in terms of the widths $\sigma_z$ and $\sigma_x$.   
Different panels refer to field orientations ($\text{a}_1$) $\phi = 0^{\circ}$, $(\text{a}_2)$ $\phi = 30^{\circ}$, $(\text{a}_3)$ $\phi = 60^{\circ}$ and $(\text{a}_4)$ $\phi = 90^{\circ}$. 
The white crosses denote the location of the respective energy minimum associated with the equilibrium set ($\sigma_x$, $\sigma_z$). 
The sign of the energy subsequently characterizes the equilibrium state as SF or self-bound. 
The equilibrium state energy between the variational method and the eGPE results are in good agreement exhibiting an average error not more than $10 \%$. ($\text{b}$) The response of (left-axis) $\sigma_{\eta}$ with $\eta \equiv\{x,y,z\}$ and (right-axis) the ratio $\sigma_y/\sigma_z$ for varying $\phi$ in the quasi-2D regime with fixed $\epsilon_{dd}=1.75$ and $N=6\times 10^4$. 
The results obtained through the variational principle (dashed lines) and the eGPE method (solid lines) are in agreement.}
\label{fig_var}
\end{center}
\end{figure}

Let us showcase that the SF and ${\rm {DL}}_{{\rm S}}$ ground-state characteristics of the dBEC phase diagram can be obtained within a variational approach instead of numerically solving the time-independent 3D eGPE. 
Particularly, our model is based on the assumption that the dBEC wave function acquires the Gaussian ansatz~\cite{Goral_2021,pal2020excitations} 
\begin{equation}
\psi(x,y,z)= \sqrt{\frac{N}{\pi^{3/2}\sigma_x \sigma_y \sigma_z}}\prod_{\eta=x,y,z} \exp(-(\frac{\eta^2}{2\sigma_\eta^2})+i\eta^2\beta_\eta(t)),\label{psivar}
\end{equation}
where the variational parameters are the widths $\sigma_\eta$ in the $\eta=x,y,z$ direction and $\beta_\eta$ which determines the phase curvature. 
It is apparent from the functional form of the ansatz that it can not capture a droplet lattice or a SS structure. 
The Lagrangian $L(\textbf{r})=\int d^3r\mathcal{L}(\textbf{r})$, with $\mathcal{L}(\textbf{r})$ being the Lagrangian density 
\begin{equation}
\begin{split}
\mathcal{L}= & \frac{i\hbar}{2}\left(\psi\frac{\partial\psi^*}{\partial t}-\psi^*\frac{\partial\psi}{\partial t}\right)+\frac{\hbar^2}{2m}\abs{\nabla\psi(\textbf{r},t)}^2
+V(\textbf{r})\abs{\psi(\textbf{r},t)}^2\\& +
\frac{g}{2}\abs{\psi(\textbf{r},t)}^4+\frac{1}{2}\int d^3r^\prime U_{dd}(\textbf{r}-\textbf{r}^\prime)\abs{\psi(\textbf{r}^\prime,t)}^2\abs{\psi(\textbf{r},t)}^2 \\&
+\frac{2}{5}\gamma_{QF}\abs{\psi(r,t)}^5.
\end{split}\label{lden}
\end{equation} 
Inserting the ansatz of Eq.~\eqref{psivar} into Eq.~\eqref{lden} and integrating over the spatial coordinates we obtain
\begin{equation}
\begin{split}\label{L_append}
L=&\sum_{\eta=x,y,z}\left[\frac{N\hbar}{2}\dot{\beta_\eta}\sigma_\eta^2+\frac{N\hbar^2}{2m}\left(\frac{1}{2\sigma_\eta^2} 
+2\beta_\eta^2\sigma_\eta^2\right)+
\frac{Nm}{4}\omega_\eta^2\sigma_\eta^2\right]\\&+\frac{gN^2}{4\sqrt{2}\pi^{3/2}}\frac{1}{\prod_{\eta} \sigma_\eta}+\frac{N^2\hbar^2}{\sqrt{2\pi}m}\frac{a_{dd}}{\prod_{\eta}\sigma_\eta}\left(\frac{3 \cos^2 \phi-1}{2} \right)  \times \\&  f(k_x,k_y)
+\frac{4\sqrt{2}\gamma_{QF}N^{5/2}}{25\sqrt{5}\pi^{9/4}}\frac{1}{\prod_{\eta}\sigma_\eta^{3/2}},	
\end{split}
\end{equation}
where the parameter $k_i=\sigma_z/\sigma_i$ ($i=x,y$) and the function 
\begin{align}\label{f_append}
f(k_x,k_y)&=\frac{1}{4\pi}\int_{0}^{\pi}d\theta \sin\theta\int_{0}^{2\pi}d\phi\nonumber\\ &\left[\frac{3\cos^2\theta}{(k_x^2\cos^2\phi+k_y^2\sin^2\phi)\sin^2\theta+\cos^2\theta}-1\right].
\end{align}
Next, by utilizing the Euler-Lagrange equations of motion for $\sigma_\eta$ and $\beta_\eta$ we arrive at the coupled set of equations
\begin{equation}
\begin{split}
&\beta_{\eta}=\frac{m}{2\hbar\sigma_{\eta}}\frac{d\sigma_{\eta}}{dt}\\&
Nm\frac{d^2\sigma_{\eta}}{d\tau^2}=-\frac{\partial}{\partial\sigma_{\eta}}U(\sigma_{\eta}).
\end{split}
\label{e-l1}	
\end{equation}
These six Euler-Lagrange equations (\ref{e-l1}) constitute exact solutions of the time-independent eGPE. 
Moreover, in Eq.~(\ref{e-l1}), the effective potential energy $U(\sigma_{\eta})$ is given by
\begin{equation}
\begin{split}
U(\sigma_{\eta})=&\sum_{\eta}\left[\frac{N\hbar^2}{m}\frac{1}{2\sigma_{\eta}^2}+\frac{Nm}{2}\omega_{\eta}^2\sigma_{\eta}^2\right]+\frac{gN^2}{2\sqrt{2}\pi^{3/2}}\frac{1}{\prod_{\eta}\sigma_{\eta}}\\&+
\sqrt{\frac{2}{\pi}}\frac{N^2\hbar^2}{m}\frac{a_{dd}}{\prod_{\eta}\sigma_{\eta}}\left(\frac{3\cos^2\phi-1}{2}\right) f(k_x,k_y)\\&+\frac{8\sqrt{2}\gamma_{QF}N^{5/2}}{25\sqrt{5}\pi^{9/4}}\frac{1}{\prod_{\eta}\sigma_{\eta}^{3/2}}. 
\end{split}\label{ueff}
\end{equation}
Apparently, the second set of Eq.~(\ref{e-l1}) is reminiscent of the classical equations of motion of a particle with coordinates $\sigma_{\eta}$ subjected to the external potential $U$. 
As such the total energy of the dBEC reads
\begin{equation}
E=\frac{N}{2}m\left(\frac{1}{2}\sum_{\eta}\left[\dot{\sigma}_{\eta}\right]^2\right)+U(\sigma_{\eta}).\label{energy_var}
\end{equation}
Therefore, the ground-state energy of the dBEC is simply $E^{(0)}=U(\sigma_{\eta}^*)$, where $\sigma_{\eta}^*$ denote the equilibrium widths. 
These are determined through minimization of the energy or equivalently the effective potential $U(\sigma_{\eta})$. 

The resulting energy $E$ as obtained from Eq.~(\ref{energy_var}) with respect to $\sigma_x=\sigma_y$, $\sigma_z$ and for various angles $\phi$ of the magnetic field is provided in Fig.~\ref{fig_var}($\text{a}_1$)-($\text{a}_4$). 
To illustrate the equivalence of the eGPE results discussed in the main text to the variational treatment we employ the quasi-2D dBEC with $\epsilon_{dd}=1.87$ and $a_{dd}=131a_B$ containing $N=6\times 10^4$ atoms in a harmonic trap with $(\omega_x, \omega_y,\omega_z)=2 \pi \times (45,45,133)$Hz. 
The equilibrium widths ($\sigma_x$, $\sigma_y$, $\sigma_z$) of the dBEC are then easily identified by determining the minimum $E^{0}$ of the energy $E$, see the white crosses in Figs.~\ref{fig_var} ($\text{a}_1$)-($\text{a}_4$). 
Interestingly, the energy minima [Figs.~\ref{fig_var}($\text{a}_1$)-($\text{a}_4$)] enable us to appreciate the phase of the dBEC that each angle $\phi$ favors. 
Indeed, we find that for $\phi=0^{\circ}$ the minimum energy is negative which is a property associated with the development of a self-bound macro droplet. 
In contrast, in the case of either $\phi = 30^{\circ}$ 
[Fig.~\ref{fig_var}($\text{a}_2$)] or $\phi = 60^{\circ}$ 
[Fig.~\ref{fig_var}($\text{a}_3$)] $E^{0}$ is positive, thus being representative of the SF phase.  
Recall that the same behavior has been concluded within the eGPE in the quasi-2D geometry [Fig.~\ref{phae_diag_pan}($\text{a}_2$)]. 
Turning to $\phi=90^{\circ}$, again the equilibrium state has negative energy [Fig.~\ref{fig_var}($\text{a}_4$)], a behavior that is related to the single droplet state discussed in Fig.~\ref{fig_quasi_2d_den2}($\text{a}_4$)--($\text{a}_8$).

As a further proof-of-principle of our benchmark we present in Figs.~\ref{fig_var} ($\text{b}$) the equilibrium widths $\sigma_x=\sigma_y\equiv \sigma_r$ and $\sigma_z$ as predicted in both the variational and the eGPE methods for the quasi-2D geometry in terms of $\phi$. It becomes evident that the equilibrium widths show almost the same behavior in both methods. 
A discernible difference is that the variational calculation overestimates (underestimates) the value of $\sigma_z$ ($\sigma_y$) until $\phi \approx 20^{\circ}$ but underestimates (overestimates) $\sigma_z$ ($\sigma_y$) for $20^{\circ}<\phi <80^{\circ}$. 
Otherwise, they agree. 
Similar conclusions can be drawn for the quasi-1D dBEC (not shown).

\section{Further details on the numerical implementation}\label{numerics}

For the convenience of our numerical simulations we cast the eGPE of Eq.~\eqref{eGPE} into a dimensionless form. 
This is achieved by rescaling the length, and time in terms of the harmonic oscillator length $l_{\rm osc} = \sqrt{\hbar/m \omega_x}$, and the trap frequency $\omega_x$ respectively, while the transformed wave function obeys $\Psi(\vb{r'}, t') = \sqrt{l^3_{\rm{osc}}/N}\psi(\vb r, t)$. 
The resulting equation is solved using the split-time Crank-Nicholson discretization scheme~\cite{crank_nicolson_1947, ANTOINE20132621}. 
The stationary (lowest energy) states of the dBEC are obtained through imaginary time propagation, effectively a gradient descent algorithm. 
At each time-step of this procedure we apply the transformation $\psi(\vb{r}', t) \rightarrow N^{1/2}/{\norm{\psi(\vb{r}', t)}}$ (for the desired $N$). 
This preserves the normalization of the wave function, while convergence is reached as long as relative deviations of the wave function (at every grid point) and energy between consecutive time-steps are smaller than $10^{-6}$ and $10^{-8}$ respectively.   
This solution is then used as an initial state for the quench dynamics where the eGPE is propagated in real time. 
Since the dipolar potential (Eq.~\eqref{avgddi}) is divergent at short distances it is calculated in momentum space, see also Ref.~\cite{Goral_2021} for the analytical expression of the Fourier transformation of the dipolar potential. 
Afterwards, we perform the inverse Fourier transform for obtaining the real space contributions using the convolution theorem. 
Our simulations are carried out in a 3D box characterized by a grid ($n_x \times n_y \times n_z$) corresponding to ($256\times256\times128$) and ($300\times600\times300$) for the quasi-2D and the quasi-1D trap respectively. 
The employed spatial discretization (grid spacing)  
refers to $\Delta x = \Delta y = \Delta z =0.1~l_{{\rm osc}}$, while the time-step of the numerical integration is $\delta t = 10^{-5}/\omega_x$.

Last but not least, our numerical approach to solve the eGPE for describing the properties of dBECs has been carefully benchmarked. 
As such, we have confirmed that it is possible to reproduce a plethora of phases appearing in the presence of a static magnetic field, e.g., from Refs.~\cite{Chomaz_2019,Blakie_2020_tube,baillie2018droplet}, as well as results where a time-averaged dipolar potential with a tilted magnetic field has been used, for instance, according to Refs.~\cite{PhysRevLett.89.130401,Prasad_2019}.
Moreover, we have meticulously checked that in our setting (where the LHY contribution is present) the time-averaged approach leads to the same results as explicitly following the time-dependent DDI. 

\begin{figure}
\includegraphics[width=0.49\textwidth]{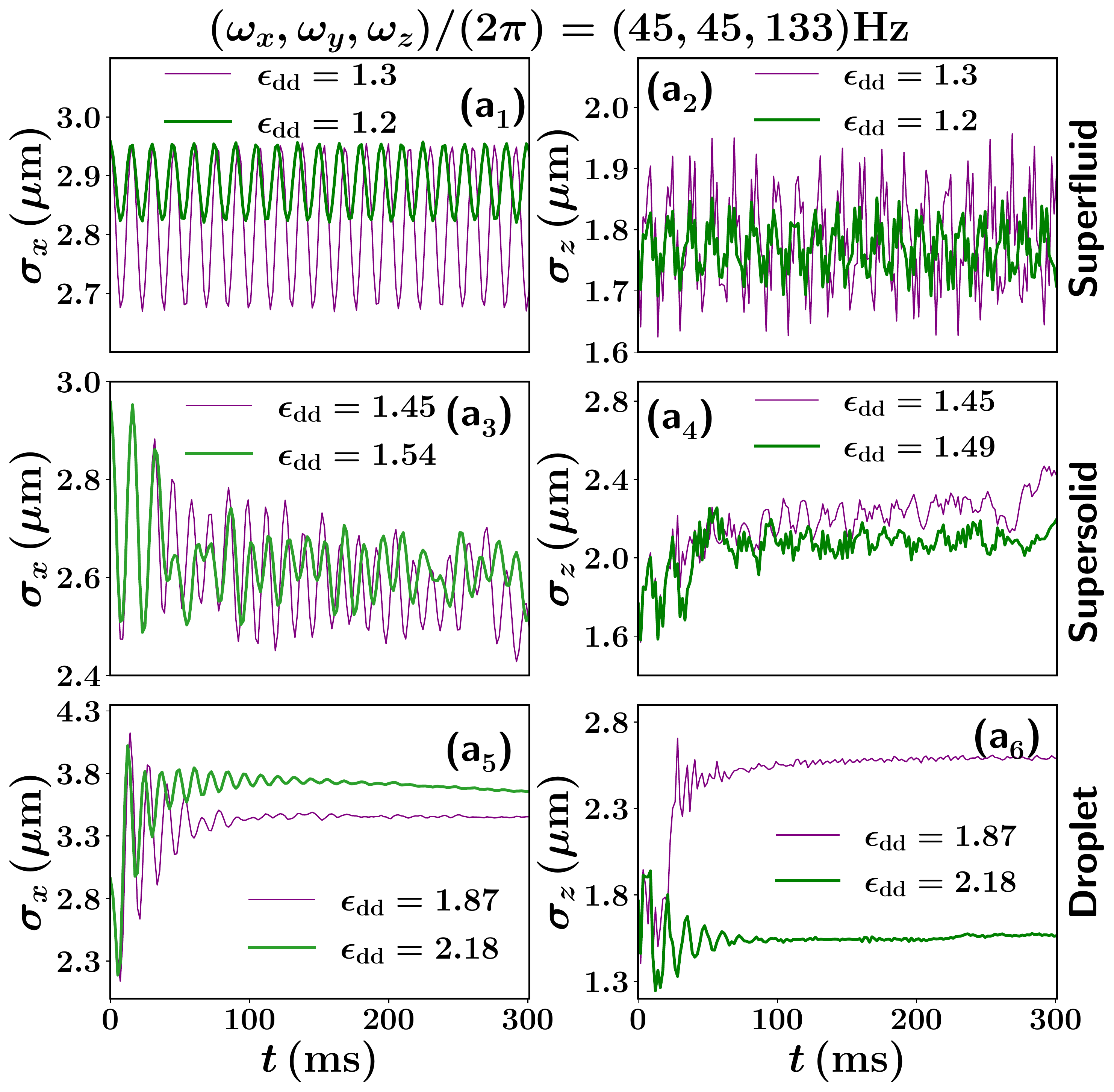}
\caption{Time-evolution of the dBEC widths [($\text{a}_1$), ($\text{a}_3$), ($\text{a}_5$)] $\sigma_x$ and [($\text{a}_2$), ($\text{a}_4$), ($\text{a}_6$)] $\sigma_z$ quantifying the breathing motion of the background caused by the quench. 
A quench from a SF state with $\epsilon_{dd}=1.1$ towards $(\text{a}_1)$-$(\text{a}_2)$ the SF, $(\text{a}_3)$-$(\text{a}_4)$ the SS and $(\text{a}_5)$-$(\text{a}_6)$ the $\rm DL_{M}$ phases characterized by specific post-quench values of $\epsilon_{dd}$ (see legend). 
Apparently, a saturation trend of the dynamically formed droplet lattice occurs [panels ($\text{a}_5$), ($\text{a}_6$)]. 
The dBEC consists of $N=6 \times 10^{4}$ atoms confined in quasi-2D trap with $(\omega_x, \omega_y, \omega_z)= 2\pi \times (45,45,133) \rm Hz$.}\label{Quench_rms_pan}
\end{figure}

\section{Collective excitations of the quenched dBEC}\label{breathing}

As already mentioned in the main text, the dBEC undergoes a collective breathing motion~\cite{Giovanni2002,Tanzi2019} originating from the interaction quench. 
A common experimentally relevant measure for estimating the amplitude and frequency of the underlying breathing is the center-of-mass variance along the different spatial directions~\cite{Giovanni2002,Kwon2021}. 
It offers a measure of the instantaneous width of the dBEC and it is defined by
\begin{align}
\sigma_{\eta} = \sqrt{\int dxdydz \abs{\psi}^2 \eta^2},\label{variance}
\end{align} 
where $\eta=\{ x,y,z\}$. 
Below, we shall analyze the dynamics in the quasi-2D regime and therefore $\sigma_x=\sigma_y\neq\sigma_z$ because $\omega_x=\omega_y\neq \omega_z$. 
However, we should note that such a breathing dynamics takes place equally  also in the quasi-1D case (not shown).

The temporal evolution of the condensate widths in the $\lambda$-th direction utilizing a quench from a SF state with $\epsilon_{dd}=1.1$ to the SF, SS and $\rm DL_{S}$ phases is illustrated in Fig.~\ref{Quench_rms_pan}($\text{a}_1$)-($\text{a}_6$). 
For a final SF state, realized here in the cases of $\epsilon_{dd}=1.19$ and $\epsilon_{dd}=1.31$, $\sigma_x(t)$ exhibits an almost constant amplitude oscillatory behavior describing the in-plane compression and expansion dynamics of the cloud [Fig.~\ref{Quench_rms_pan}($\text{a}_1$)]. 
As expected, the oscillation (breathing) amplitude is reduced for smaller quench amplitudes. For the respective frequency we find that $\sigma_x(t)$ oscillates in-phase with $\omega^{x(y)}_{\rm SF} \approx 67 \rm Hz$ at the early stages of the evolution for both $\epsilon_{dd}=1.19$ and $\epsilon_{dd}=1.31$. 
Later on, $\sigma_x(t)$ possesses a smaller frequency $\omega^{x(y)}_{\rm SF} \approx 64 \rm Hz$ for $\epsilon_{dd}=1.31$. 
Unlike $\sigma_x(t)$, in the transversal direction $\sigma_z(t)$ exhibits multifrequency oscillations of time varying amplitude [Fig.~\ref{Quench_rms_pan}($\text{a}_2$)]. 
Generally, the breathing motion does not decay in the SF regime. 

Turning to a SS post-quench state we observe that $\sigma_x(t)$ experiences a peculiar beating pattern characterized by two dominant frequencies [Fig.~\ref{Quench_rms_pan}($\text{a}_3$)]. 
They correspond to $\omega^{x(y)}_{\rm SS,1} \approx 73.147 \rm Hz$ and $\omega^{x(y)}_{\rm SS, 2} \approx 60.8 \rm Hz$ for $\epsilon_{dd}=1.45$, while $\omega^{x(y)}_{\rm SS,1} =75.96 \rm Hz$ and $\omega^{x(y)}_{\rm SS,2} \approx 17.5 \rm Hz$ for $\epsilon_{dd}=1.53$. The mode of higher frequency is related with the deformation of the SS lattice, and the lower one to the collective motion of the background superfluid, see also Fig.~\ref{den_pan_quench}(a$_1$)-(a$_6$). 
Evidently, upon reducing $a_s$, the higher (lower) frequency mode increases (decreases). The involvement of these low frequency compressional (breathing) modes is inherently related to the manifestation of the supersolid state, see also Ref.~\cite{Tanzi_2019}.
This response is anticipated since for a smaller $a_s$ the background density is more dilute, and the crystal structure becomes more prominent. 
As such, the compressional mode associated with the crystal hardens whilst the lower one vanishes~\cite{Tanzi2019}. 
Concluding, $\sigma_z(t)$ initially features an increase while fluctuating and after the formation of the SS lattice it shows a saturation tendency [Fig.~\ref{Quench_rms_pan}($\text{a}_4$)].

For the droplet region, a completely different response takes place [Fig.~\ref{Quench_rms_pan}($\text{a}_5$)-($\text{a}_6$)]. 
Particularly, $\sigma_x(t)$ initially increases and after the formation of the droplet cluster around $t =4 \rm ms$ [see also Fig.~\ref{den_pan_quench}($\text{b}_4$)], it shows an oscillatory trend of decaying amplitude [Fig.~\ref{Quench_rms_pan}($\text{a}_5$)], signaling the  collective expansion and contraction of the lattice. 
Afterwards, in the long-time dynamics, $\sigma_x(t)$ saturates 
capturing the stationary configuration of the cluster. 
A similar response can be seen in $\sigma_z(t)$ [Fig.~\ref{Quench_rms_pan}($\text{a}_6$)]; however, unlike $\sigma_x(t)$, the growth of $\sigma_z(t)$ is larger for increasing post-quench $a_s$ implying a larger amount of transversal excitations. 

\bibliographystyle{apsrev4-1}
\bibliography{reference}	
\end{document}